%% file: ms.tex
\documentclass[conference]{IEEEtran}
\IEEEoverridecommandlockouts
\usepackage{cite}
\usepackage{amsmath,amssymb,amsfonts}
\usepackage{algorithmic}
\usepackage{graphicx}
\usepackage{textcomp}
\usepackage{xcolor}

\input{physcomplab-preamble.tex}

\def\BibTeX{{\rm B\kern-.05em{\sc i\kern-.025em b}\kern-.08em
    T\kern-.1667em\lower.7ex\hbox{E}\kern-.125emX}}
\begin{document}
	
\title{A System for Generating Non-Uniform Random Variates using Graphene Field-Effect Transistors}

\author{\IEEEauthorblockN{Nathaniel J. Tye}
\IEEEauthorblockA{\textit{Cambridge Graphene Centre,} \\
\textit{Department of Engineering,} \\ 
\textit{University of Cambridge}\\
njt48@cam.ac.uk}
\and
\IEEEauthorblockN{James T. Meech}
\IEEEauthorblockA{\textit{Department of Engineering} \\
\textit{University of Cambridge}\\
jtm45@cam.ac.uk}
\and
\IEEEauthorblockN{Bilgesu A. Bilgin}
\IEEEauthorblockA{\textit{Department of Engineering} \\
\textit{University of Cambridge}\\
bab46@cam.ac.uk}
\and
\IEEEauthorblockN{Phillip Stanley-Marbell}
\IEEEauthorblockA{\textit{Department of Engineering} \\
\textit{University of Cambridge}\\
phillip.stanley-marbell@eng.cam.ac.uk}
}

\maketitle

\input{abstract.tex}


\input{introduction.tex}

\input{gfet-overview.tex}

\input{gfet-prva.tex}
\input{gfet-fabrication-methodology.tex}

\input{gfet-simulation-results.tex}

\input{monte-carlo-results.tex}

\input{relatedwork.tex}
\input{conclusions.tex}

 \section*{Acknowledgements}
 This research is supported by an Alan Turing Institute award TU/B/000096 
under EPSRC grant EP/N510129/1, by EPSRC grant EP/R022534/1, and by 
EPSRC grant EP/V004654/1. N.J. Tye acknowledges funding from EPSRC 
grant EP/L016087/1. J.T. Meech acknowledges funding from EPSRC grant
EP/L015889/1. We acknowledge contributions from Stephan Hofmann,
Jack Alexander-Webber and Ye Fan, as well as funding from Huawei Finland.

\bibliographystyle{abbrv}
\bibliography{ms}

\end{document}

%% file: physcomplab-preamble.tex
\usepackage{framed}
\usepackage{moreverb}
\usepackage[small,sf,bf]{subfigure}
\usepackage{xspace}
\usepackage{flushend}			%
\usepackage{siunitx}
\usepackage{xfrac}			%

\usepackage[ruled,vlined]{algorithm2e}

\usepackage{amsfonts}
\usepackage{amssymb}
\usepackage{amsmath}
\usepackage[thmmarks,amsmath,hyperref]{ntheorem}
\usepackage{graphicx}
\usepackage{times}
\usepackage[scaled]{helvet} 		%
\usepackage{textcomp}
\usepackage{colortbl}
\usepackage{color}
\usepackage{booktabs}
\usepackage{listings}
\usepackage{multicol}
\usepackage{times}
\usepackage{pifont}			%
\usepackage{microtype}

\usepackage[scaled=0.85]{beramono}
\usepackage[T1]{fontenc}
\usepackage{url}
\usepackage{multirow}
\usepackage{array}

\definecolor{listinggreen}{rgb}{0,0.6,0}
\definecolor{listinggray}{rgb}{0.5,0.5,0.5}
\definecolor{listingmauve}{rgb}{0.58,0,0.82}
\definecolor{listingkeywordcolor}{rgb}{1.0,0.4,0.0}
\definecolor{listinglightgray}{rgb}{0.9863,0.9863,0.9863}
\definecolor{a}{rgb}{0.9,0.95,0.95}%
\definecolor{b}{rgb}{0.99,0.99,0.99}%

\lstdefinelanguage{FSharp}
{
	morekeywords	= {
    let,
    type,
    Measure,
	},
	sensitive	= false,
	morecomment	= [l]{\#},
	morecomment	= [s]{(*}{*)},
}

\lstdefinelanguage{Newton}
{
	morekeywords	= {
		signal,
		derivation,
		symbol,
		name,
		invariant,
		constant,
		English,
		sensor,
		name,
		none,
		dot,
		cross,
		derivative,
		integral,
		interface,
		i2c,
		spi,
		analog,
		write,
		read,
		delay,
		range,
		erasuretoken,
		uncertainty,
		accuracy,
		precision,
		Gaussian,
		exponential,
		biexponential,
		to,
		bits,
		dimensionless,
		include
	},
	sensitive	= false,
	morecomment	= [l]{\#},
	morecomment	= [s]{/*}{*/},
}

\usepackage{hyperref}
\hypersetup{
    colorlinks=false, %
    linktoc=all,     %
    linkcolor=blue,  %
}

%% file: abstract.tex
\begin{abstract}
We introduce a new method for hardware non-uniform random number
generation based on the transfer characteristics of graphene field-effect
transistors (GFETs) which requires as few as two transistors and
a resistor. 

\quad We implement the method by fabricating multiple GFETs and
experimentally validating that their transfer characteristics exhibit
the nonlinearity on which our method depends.
We use characterisation data in simulations of a proposed
architecture for generating samples from dynamically selectable
non-uniform probability distributions. The method we present has the potential 
for Gb/s sample rates, is reconfigurable for arbitrary target distributions, 
and has a wide range of possible applications.

\quad Using a combination of experimental measurements of GFETs
under a range of biasing conditions and simulation of the GFET-based
non-uniform random variate generator, we demonstrate
a speedup of Monte Carlo integration by up to 2$\times$.
This speedup assumes the analog-to-digital converters reading the outputs
from the circuit can produce samples in the same amount of time that
it takes to perform memory accesses.
\end{abstract}

%% file: introduction.tex
\section{Introduction}
\label{section:introduction}
\vspace{-0.05in}
Hardware \textit{uniform random number generators} exist
in both research and commercial computer architectures, with
generation rates of up to 6.4 Gb/s~\cite{IntelHardware}. Uniform
random numbers are widely used in applications such as cryptography,
where the objective is to generate bit vectors (e.g., 256-bit
vectors) that are uniformly distributed over some range and are
therefore difficult to guess. In contrast, this article focuses on
\textit{programmable random variate accelerators} (PRVAs) 
for generating \textit{non-uniform
random variates} (NRVs), random samples chosen from a non-uniform
probability distribution.

\subsection{Applications of Non-Uniform Random Variates}
\vspace{-0.05in}
Many important applications in science and engineering depend not
on uniform random samples, but instead require NRVs,
from a wide range of distributions.
Applications of NRVs range
from Monte Carlo simulations~\cite{metropolis1949monte},
to quantitative finance~\cite{chen2017monte} to particle filter localisation
for driverless cars~\cite{thrun2010toward}. NRVs are also important in Bayesian machine learning
applications~\cite{lambert2018student}, which involve the
computation of a marginal probability which goes into the denominator
of the expression of Bayes' rule. Computing these marginal
probabilities in turn requires evaluating an integral of a probability
distribution.  Because the distributions in question are typically
high-dimensional and have no known analytic equational form, their
integration often requires Monte Carlo integration methods
where one samples repeatedly from the corresponding distribution.

\subsection{Challenges}
\vspace{-0.05in}
Because generating samples from distributions whose inverse cumulative
distribution function (CDF) does not have a closed form requires time-consuming accept-reject sampling~\cite{devroye2008non}, generating
random samples from non-uniform distributions is typically an order of
magnitude slower and less energy-efficient than generating uniformly
distributed random samples~\cite{thomas2009comparison}. One
promising direction for efficiently generating NRVs is to sample from a physical process whose evolution in
time~\cite{zhang2018architecting}
or noise characteristics~\cite{meech2020efficient} follow some known and
(ideally) controllable probability distribution.

\subsection{Contributions}
\vspace{-0.05in}
This article presents the first demonstration of generating NRVs by exploiting properties of graphene field-effect transistors (GFETs)
previously considered undesirable: their ambipolar transfer characteristics and their 
lack of a band gap. We provide a tutorial overview of the properties of GFETs
(Section~\ref{section:GFET-overview}) and introduce a circuit
topology for using a chain of GFETs together with a uniform random
variate generator to generate dynamically controllable non-uniform
distributions (Section~\ref{section:GFET-prva}).  We present the
methodology we used for fabricating an array of GFETs and empirically
characterising their transfer characteristics
(Section~\ref{section:GFET-fabrication}) and we use those
empirically measured GFET transfer characteristics to demonstrate
the proposed method in a simulated combined circuit topology
(Section~\ref{section:gfet-simulation-results}).  We then use the
generated NRVs in an end-to-end system
example, where we evaluate their benefit to speeding up Monte Carlo
integration, as well as their benefit to reducing the error in the
Monte Carlo integration process
(Section~\ref{section:Monte-Carlo-results}). 

%% file: gfet-overview.tex
\begin{figure}[t]
	\centering
	\includegraphics[trim=0cm 0.1cm 0cm 0.1cm, clip=true, width=0.5\textwidth]{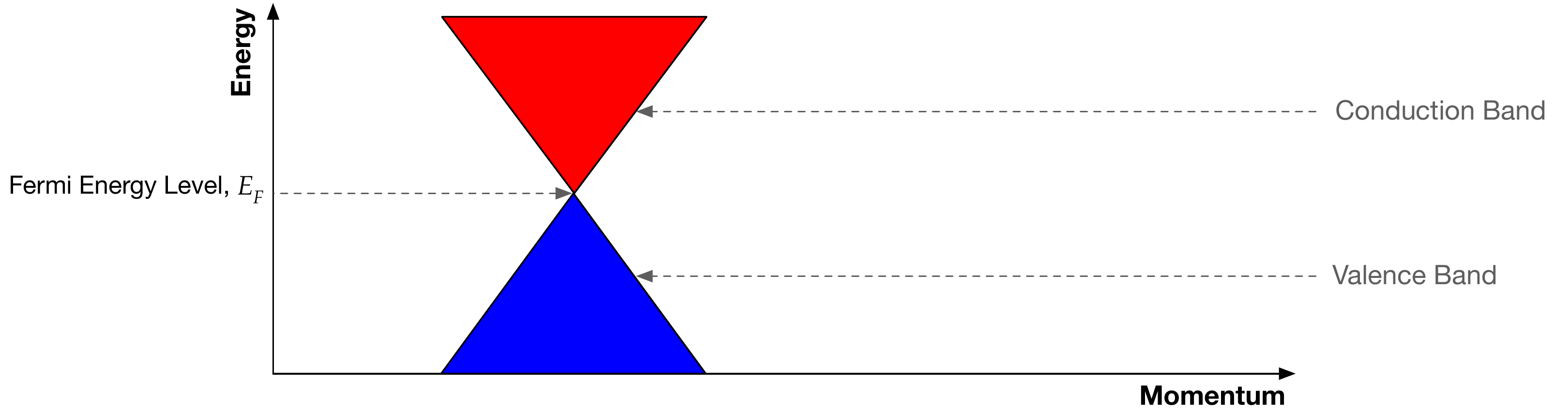}
	\vspace{-0.3in}
	\caption{Energy band structure of graphene, showing the
		Dirac point, where the conduction and valence bands touch.
		$E_F$ is the Fermi level. In undoped/unbiased graphene, it
		is located at the Dirac point. There is no band gap:
		GFETs have low on- to off-current ratios making them a
		poor choice for traditional digital logic applications~\cite{keyes1985makes}.}
	\label{figure:DiracPoint}
	\vspace{-0.25in}
\end{figure}

\section{Properties of Graphene Field-Effect Transistors}
\label{section:GFET-overview}
\vspace{-0.05in}
GFETs have a channel
made of single- or multi-layer graphene, rather than a semiconducting
material such as silicon or germanium~\cite{Schwierz2010}.
Unlike traditional semiconductors, graphene is a semi-metal:
it lacks a band gap and its conduction and valence bands don't
overlap. Instead, the conduction and valence bands meet at
the \textit{Dirac point} (Figure~\ref{figure:DiracPoint}).

Electrons at the Dirac point are effectively massless and so have
unusually high electron mobilities. As a result, the phonon-limited
carrier mobility (the highest possible mobility limited by
interactions between carriers and vibrations of the channel's crystal
lattice) of graphene on SiO$_2$ is predicted to be as high as 200,000\,$\mbox{cm}^2\mbox{V}^{-1}\mbox{s}^{-1}$~\cite{Chen2008}.
Although high electron mobilities result in more efficient flow of charge,
the lack of a band gap means GFETs have low on- to off-current ratios
and can never completely turn off, making them unsuitable for
digital logic applications~\cite{keyes1985makes}.

The poor on- to off-current ratios of GFETs in digital logic applications
does not preclude their use in other areas of computing. Because it is possible
to tune the Fermi level in graphene (which typically lies at the
Dirac point) by biasing the channel, it is possible to control
device characteristics, e.g., using multi-gate structures, in a manner not equally
possible in typical metal-oxide-semiconductor field-effect transistors
(MOSFETs). GFETs also have transfer characteristics that are non-linear and different
to MOSFETs: as the gate voltage is swept, the drain current exhibits a v-shaped 
characteristic curve, with the conductance decreasing until it reaches a minimum value 
before increasing again.

%% file: gfet-prva.tex
\section{A GFET Non-Uniform Random Variate Generator}
\label{section:GFET-prva}
\vspace{-0.05in}

If the signal at the gate of a GFET is a uniform random voltage
distribution, then the distribution of the drain current will be
modified by the GFET's transfer characteristics. The exact shape
of the transfer characteristics varies with the source-drain voltage
$V_{DS}$. Thus, for a uniform random voltage distribution
at the gate, varying $V_{DS}$ for a GFET changes the
distribution of drain currents. If these drain currents are converted
to a voltage and applied to additional GFETs, it is
possible to combine the transfer characteristics and biasing of
multiple GFETs to achieve a range of drain current (and hence
voltage) distributions.

\begin{figure}[t]
	\centering
	\includegraphics[trim=0cm 0cm 0cm 0cm, clip=true, width=0.5\textwidth]{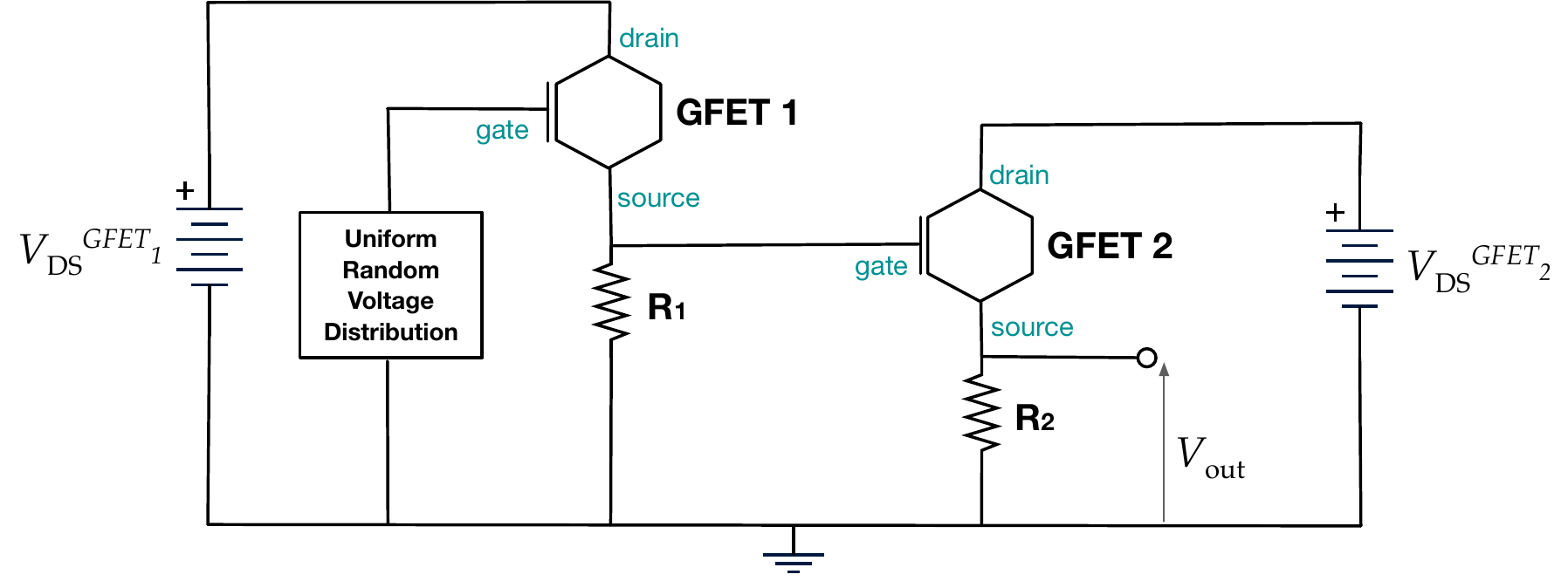}
	\vspace{-0.3in}
	\caption{Example schematic of a possible circuit used to
		transform a random uniform noise distribution (V1) into an
		arbitrary distribution by cascading several individually biased GFETs.}
	\label{figure:CircuitEg}
	\vspace{-0.15in}
\end{figure}

Figure~\ref{figure:CircuitEg} shows a possible circuit
to implement generation of a controllable non-uniform
voltage distribution using GFET properties. Each GFET in
Figure~\ref{figure:CircuitEg} has a bias voltage, $V_{DS}$,
that controls its transfer characteristics. The first GFET
has a uniformly distributed random voltage across its gate
and a corresponding distribution of drain currents, with
the values of the drain currents for each input voltage
determined by the transfer characteristics of the
GFET at its bias voltage $V_{DS}^{\mbox{\tiny GFET$_1$}}$.  The
circuit in Figure~\ref{figure:CircuitEg} converts the drain current
of the first GFET into a voltage input to the gate of the
second GFET, using a resistor, $R_1$. In practice, using a transimpedance
amplifier (TIA) to perform this current-to-voltage conversion would
result in less Johnson-Nyquist noise in the generated voltages,
though the presence of such noise may not be detrimental given our
goal of generating random variates. The analyses that follow
in Section~\ref{section:gfet-simulation-results} therefore use a
resistor for converting the drain currents to voltages to control
the second stage in the circuit. The second GFET in
Figure~\ref{figure:CircuitEg}, operating at a bias voltage of
$V_{DS}^{\mbox{\tiny GFET$_2$}}$, further shapes the distribution
of the output signal. By selectively connecting multiple GFETs
in the manner of Figure~\ref{figure:CircuitEg} (and
possibly using multi-gate GFETs), this method in principle permits
generation of a final output $V_{\mbox{out}}$ with a range of
selectable distributions, controlled by the combination of $R_1$,
$V_{DS}^{\mbox{\tiny GFET$_1$}}$, and $V_{DS}^{\mbox{\tiny GFET$_2$}}$.

\begin{figure}
	\centering
	\includegraphics[trim=0cm 0cm 0cm 0.1cm, clip=true, width=0.4\textwidth]{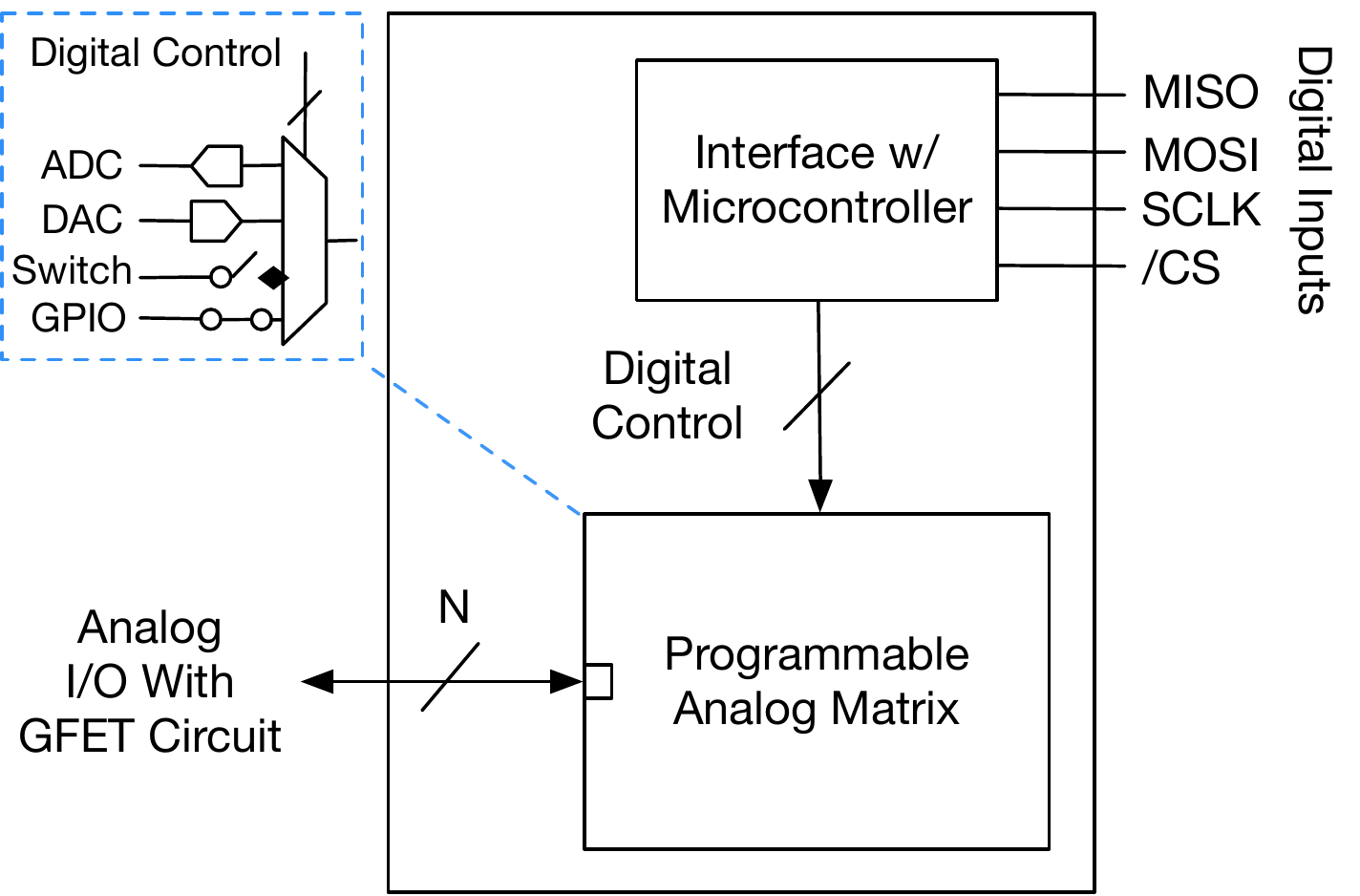}
	\vspace{-0.1in}
	\caption{Schematic of integration hardware used with the GFET circuits,
		based on the schematic of the Maxim MAX11300 \cite{MAX11300Datasheet}.
		The $N$ signal analog I/O connects to each terminal of each GFET. The zoomed
		view shows circuitry within the programmable analog matrix for control
		of each terminal. The switch determines whether the channel is enabled,
		the ADC converts an analog signal into a digital signal read by the CPU,
		the DAC converts a digital signal from the CPU to an analog signal for the GFETs. 
		The GPIO ports function as inputs/outputs with a controllable logic level, e.g.
		for setting a bias level. All of these are run into a MUX which is digitally
		controlled by the microcontroller.}
	\label{figure:MAX11}
	\vspace{-0.25in}
\end{figure}

\subsection{Integration with an  Existing Computer Architecture}

For integration into a larger system, we propose the use of a programmable
analog switching matrix, e.g., a MAX11300 \cite{MAX11300Datasheet}, such as
that illustrated in (Figure~\ref{figure:MAX11}).
This serves as an interface between an external processor
and the GFET die and allows for dynamic reconfiguration of the GFET circuit.
We present a hardware prototype of this analog switching matrix in
Section~\ref{section:GFET-fabrication}.

\begin{figure}
	\centering
	\includegraphics[width=0.48\textwidth]{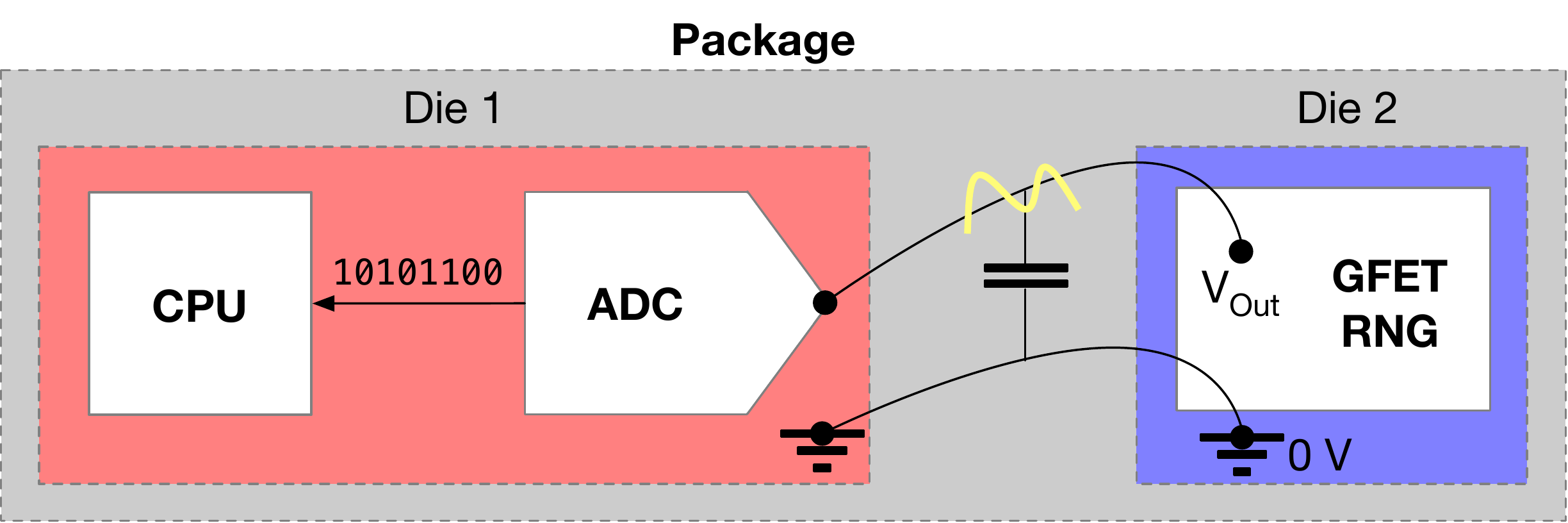}
	\vspace{-0.1in}
	\caption{Connection of CPU and random number generator die using bond wires with resistance $R$ across them and capacitance $C$ between them.}
	\vspace{-0.23in}
	\label{figure:package}
\end{figure}

The GFET random number generator could be integrated into a package with an existing
CPU and ADC using bond wires to connect the two separate dies.
Figure~\ref{figure:package} shows a block diagram of the arrangement. 
The maximum frequency at which the bond wire can change state depends on the $RC$ time constant. 

\subsection{Wavelet Decomposition and Reconstruction of Distributions}

In signal processing, the Fourier transform decomposes a time-varying signal
into its constituent frequency components and the Fourier series allows for the
construction of an arbitrary signal from a sum of sines and cosines. This is a
special case of wavelet analysis, which allows for any function to be described
by a set of orthonormal basis functions.

Wavelet analysis uses an analysing wavelet with a scaling function to
generate a set of basis functions. These basis functions are simply scaled and
shifted versions of the analysing wavelet \cite{Graps1995}. The inner product
of the scaling and wavelet functions, which are necessarily orthogonal, gives a
matrix of wavelet coefficients.

The discrete wavelet transform (DWT) uses known scaling and wavelet functions
to generate a known set of basis functions. When applied to a discrete signal,
those basis functions give an approximation of the signal, characterised by its 
wavelet coefficients \cite{Daubechies1992}. The inverse
transform is simply the linear combination of these basis functions,
and thus allows for reconstruction of the original signal with a desired level of
accuracy dependent on the number of coefficients used.

\begin{figure}[h]
	\centering
	\subfigure[]{\includegraphics[trim=2cm 8.1cm 2cm 8.5cm, clip=true, width=0.49\linewidth]{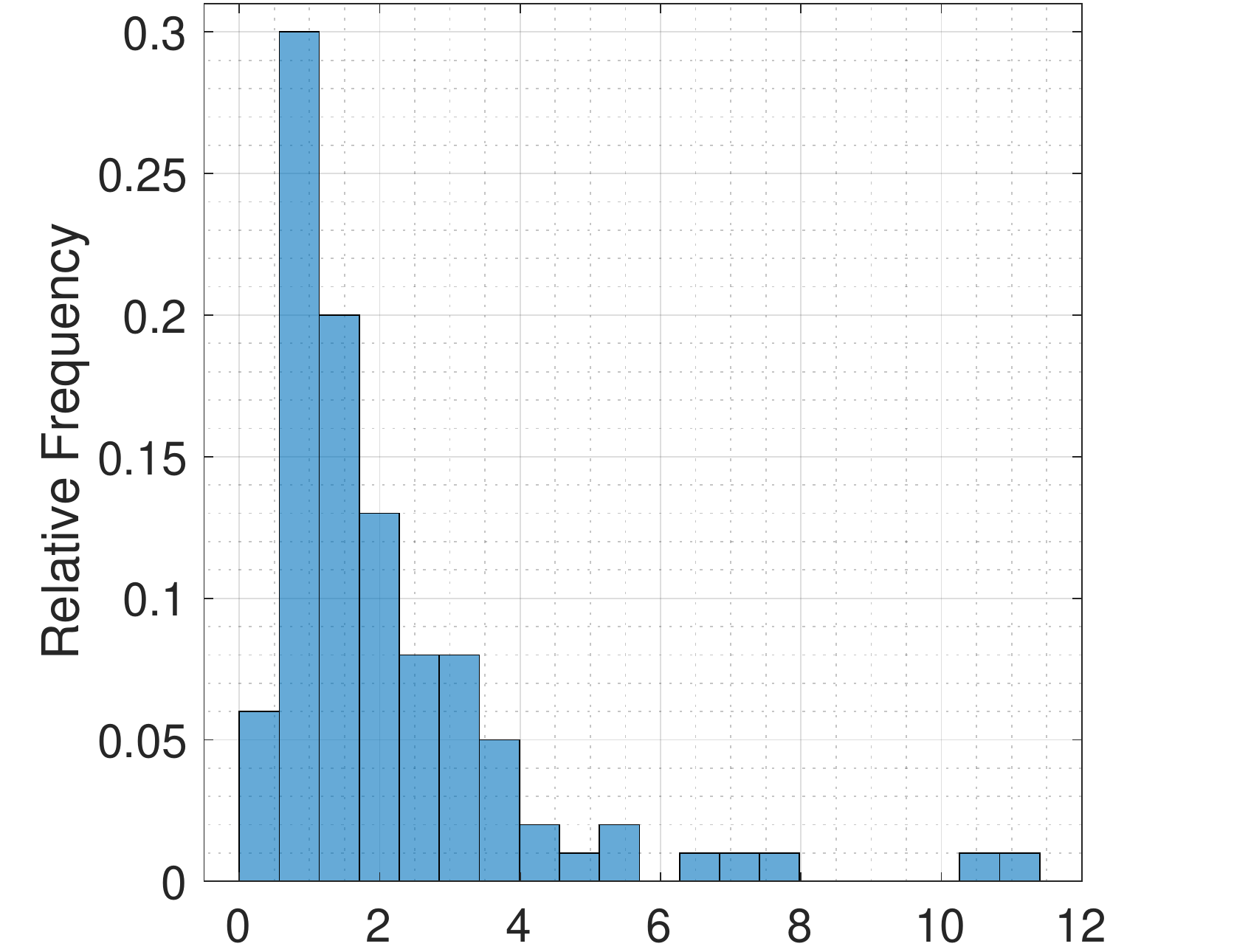}}
	\subfigure[]{\includegraphics[trim=2cm 8.1cm 2cm 8.5cm, clip=true, width=0.49\linewidth]{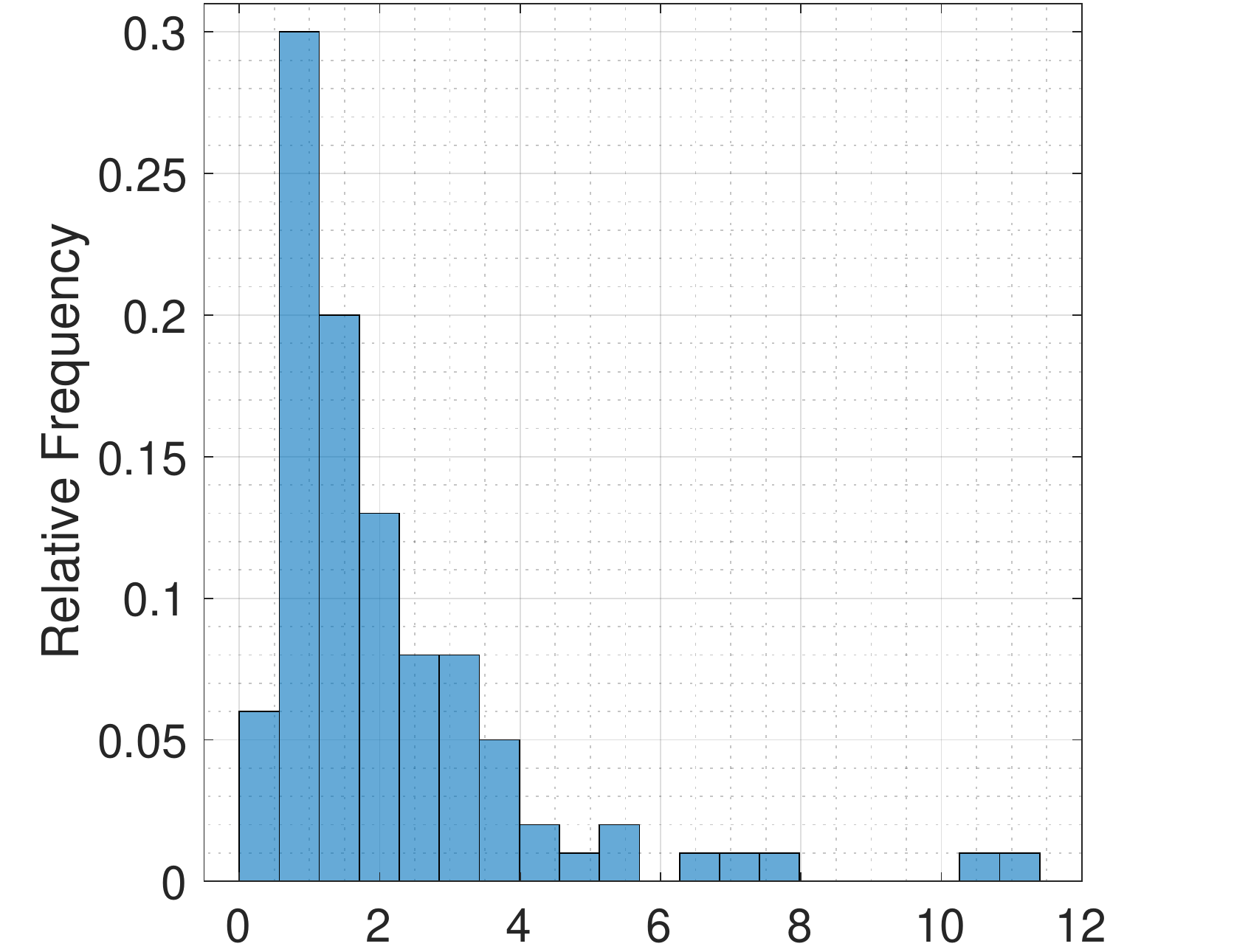}}\\[-0.14in]
	\subfigure[]{\includegraphics[trim=2cm 8.1cm 2cm 8cm, clip=true, width=0.49\linewidth]{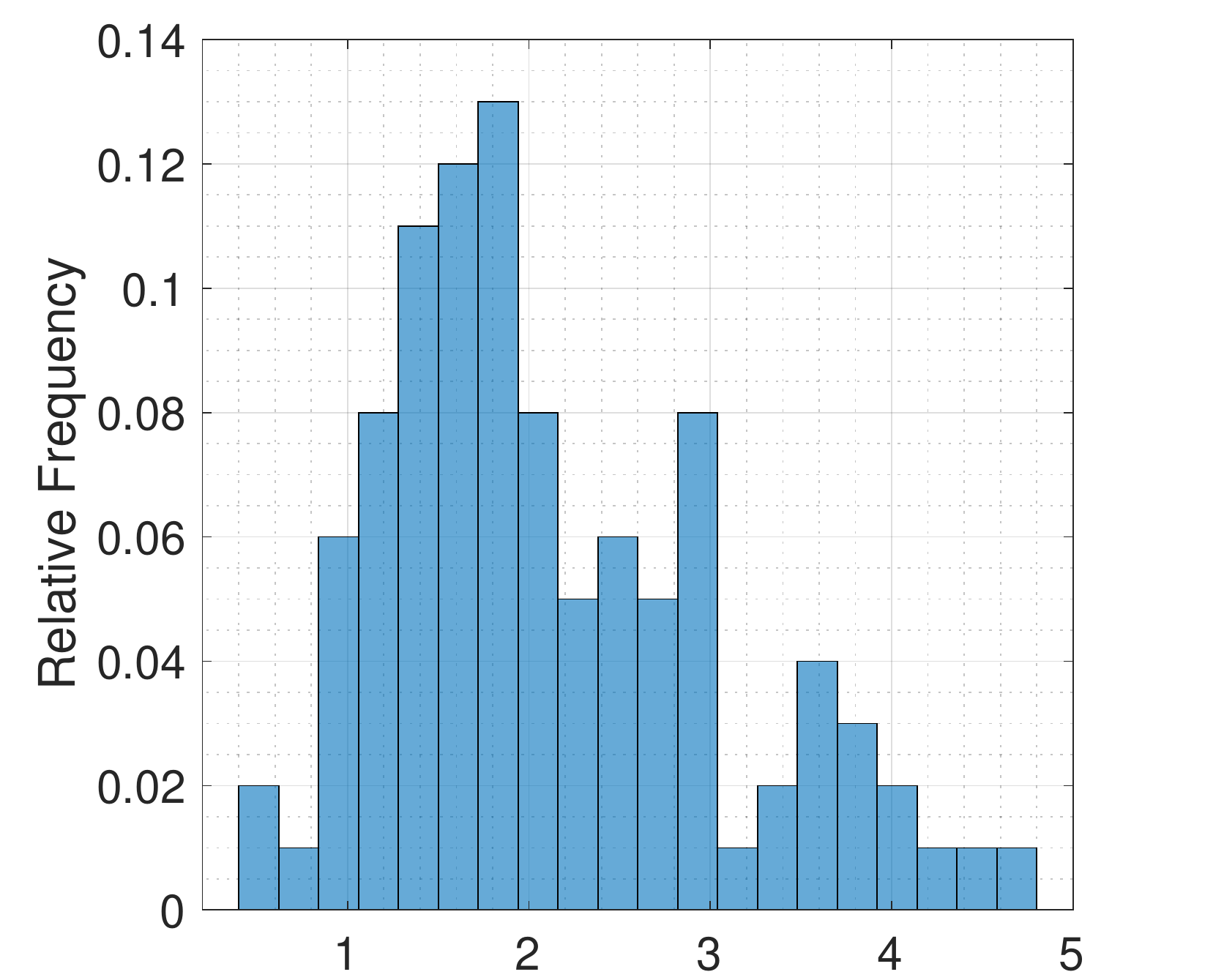}}
	\subfigure[]{\includegraphics[trim=2cm 8.1cm 2cm 8cm, clip=true, width=0.49\linewidth]{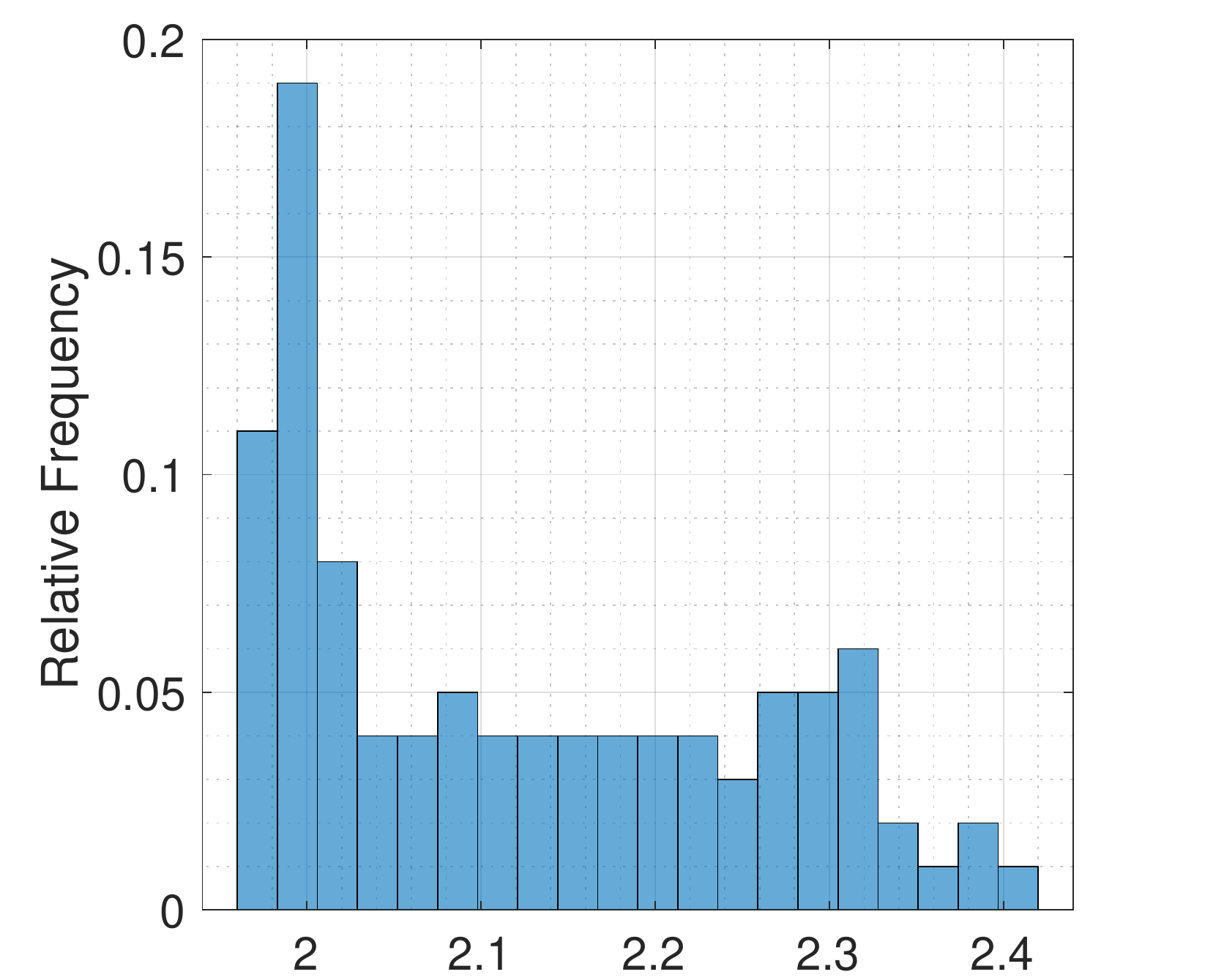}}
	\vspace{-0.15in}
	\caption{\textbf{(a)} A software-generated lognormal distribution;
		\textbf{(b)} reconstructed distribution using 55 coefficients;
		\textbf{(c)} reconstruction using 22 coefficients;
		\textbf{(d)} reconstruction using 12 coefficients. In each case, the
		x-axis is simply a number.}
	\vspace{-0.23in}
	\label{figure:WaveletExample}
\end{figure}

\begin{figure}[h]
	\centering
	\includegraphics[trim=0cm 0cm 0cm 0cm, clip=true, width=0.395\textwidth]{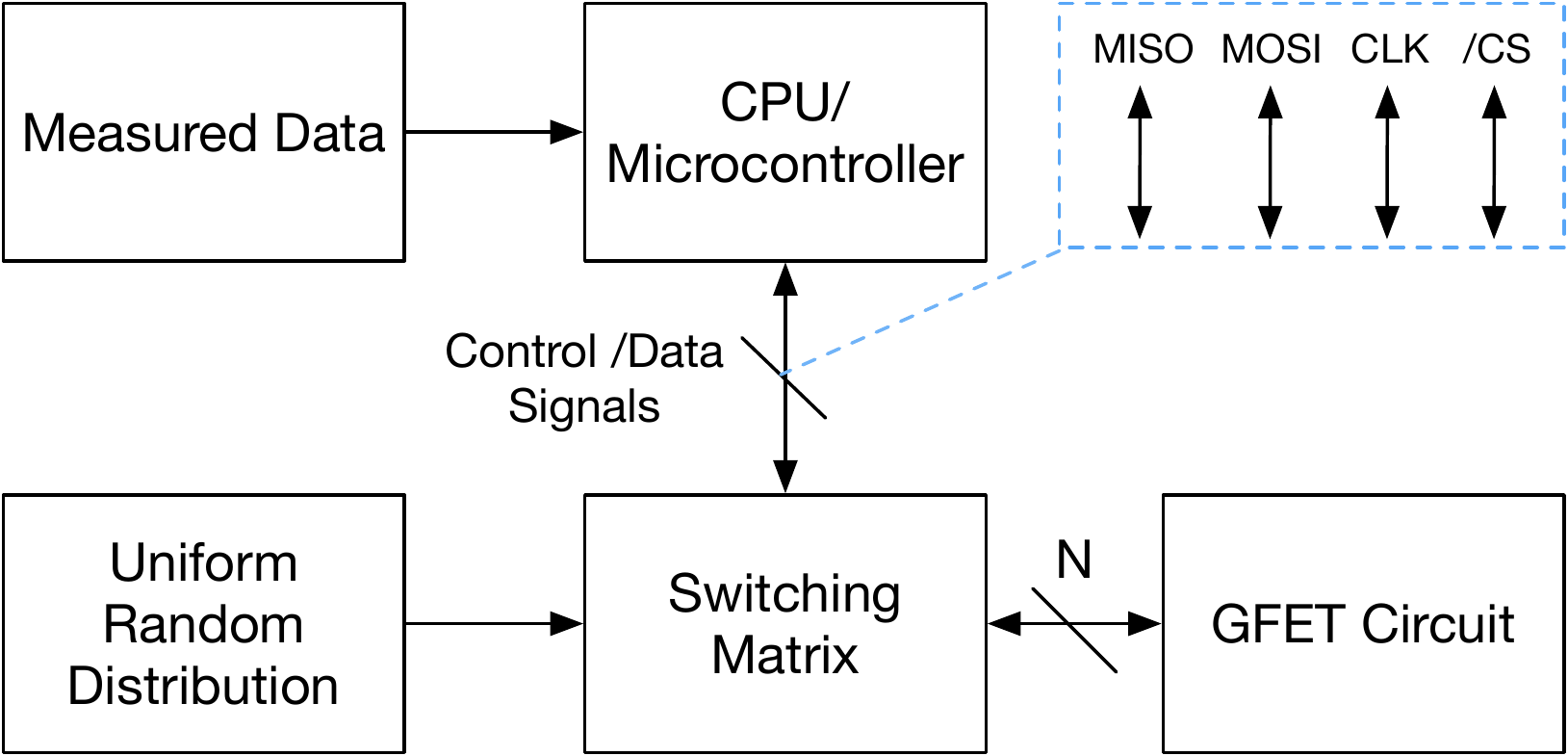}
	\vspace{-0.05in}
	\caption{Proposed system to sample from non-uniform distributions. A CPU
		processes some data and takes the DWT, then converts the DWTs scaling
		parameters into control signals, e.g., bias voltages and configurations of GFETs.
		The architecture inputs a uniform random variate to the GFET circuit, which
		transforms it into an approximation of the target distribution. The
		CPU then reads the approximated distribution from the circuit output.}
	\label{figure:WaveletsSystem}
	\vspace{-0.2in}
\end{figure}

We demonstrate wavelet decomposition and reconstruction of a distribution in
the following example. We show the distributions reconstructed from inverse
DWTs with different numbers of coefficients, corresponding to a given level
of accuracy for the lognormal distribution in Figure~\ref{figure:WaveletExample}.
We used a second-order Coiflet wavelet for both the DWT and the inverse-DWT.
We chose this wavelet arbitrarily as a proof-of-concept for the proposed method,
but it appeared to give reasonable results. We calculated the Kullback-Leibler (KL)
divergence \cite{kullback1951}, a measure of the closeness of two distributions. Figure~\ref{figure:WaveletExample}(b) uses the most
coefficients and was thus the most accurate reconstruction, with a calculated KL
divergence of 0, an identical reconstruction. Figure~\ref{figure:WaveletExample}(c)
used less than half the coefficients of (b) and had a KL divergence of 1.19.
Figure~\ref{figure:WaveletExample}(d), which used less than a quarter of
the coefficients of (b) was actually closer than (c), with a KL divergence of 0.45. Figure~\ref{figure:WaveletsSystem} is a block diagram of a proposed complete system.
Section~\ref{section:GFET-fabrication} and Figure~\ref{figure:GFETPhoto}
(c) present an early prototype of this system.

We have shown that a distribution can be reconstructed from a set of
wavelet coefficients determined by a wavelet transform. If we consider
these coefficients to be bias voltages for GFETs, which have a tunable
characteristic transfer function with a certain distribution, then the
characteristic of a GFET can be considered as an analysing wavelet, with the
bias voltages being the scaling parameters. Summing the output of each
GFET (or combination of GFETs), with each representing a basis function,
in principle, allows for the reconstruction of any arbitrary function, with
the accuracy dependent on the number of GFETs used.

%% file: gfet-fabrication-methodology.tex
\section{GFET Fabrication and Electrical Characterisation}
\label{section:GFET-fabrication}
\vspace{-0.05in}
GFETs in the results presented here consist of an Si substrate onto
which we patterned a gold back gate. We grow a 60\,nm alumina
(Al$_2$O$_3$) layer by atomic layer deposition (ALD), onto which
we transfer a monolayer of graphene using a wet transfer process.
This graphene monolayer was grown by chemical vapour deposition (CVD)
and purchased from Graphenea Inc~\cite{Graphenea}. After patterning the graphene, we
deposit the gold source and drain contacts onto the graphene to create a
$40\,\mathrm{\mu m}\times40\,\mathrm{\mu m}$ GFET channel that lies
exactly above the back gate. The large feature size compared to GFETs in
the literature (channel lengths below 100\,nm~\cite{Liao2010}) is for ease
of fabrication and testing. The electrical performance is beyond the scope
of this work; tunability and properties of GFETs are the focus here.
The channel is insulated by another layer of alumina, onto which we pattern
gold top gates aligned with the GFET channels. We electrically passivate the
whole device using a final ALD deposition of alumina. Finally, to facilitate electrical
access to the GFETs, we etch away the alumina on top of the contacts that
are electrically linked to the source, drain, back, and top gates of the GFETs.

\begin{figure}[t]
	\centering
	\subfigure[]{\includegraphics[width=0.236\textwidth]{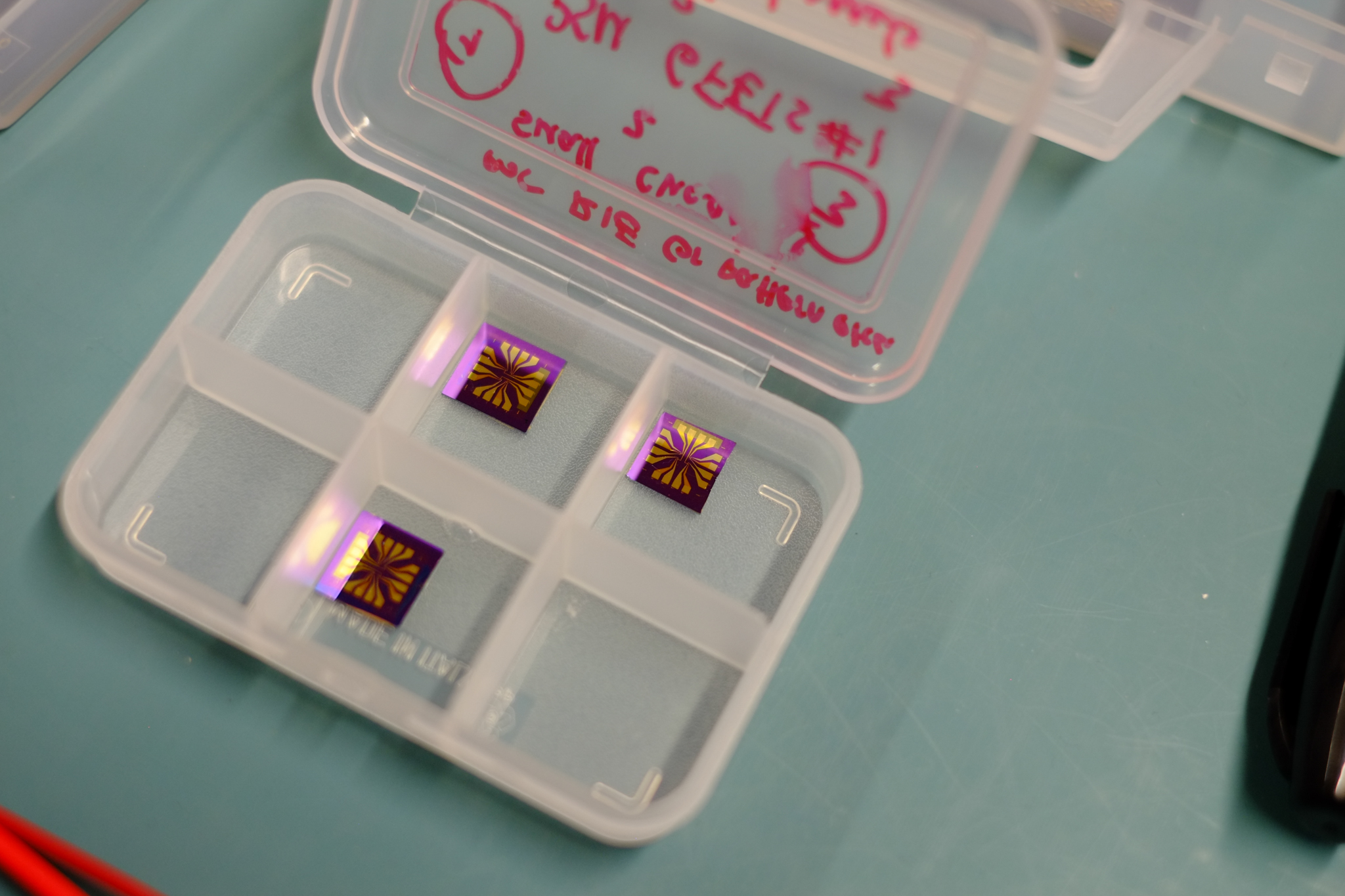}}
	\subfigure[]{\includegraphics[width=0.21\textwidth]{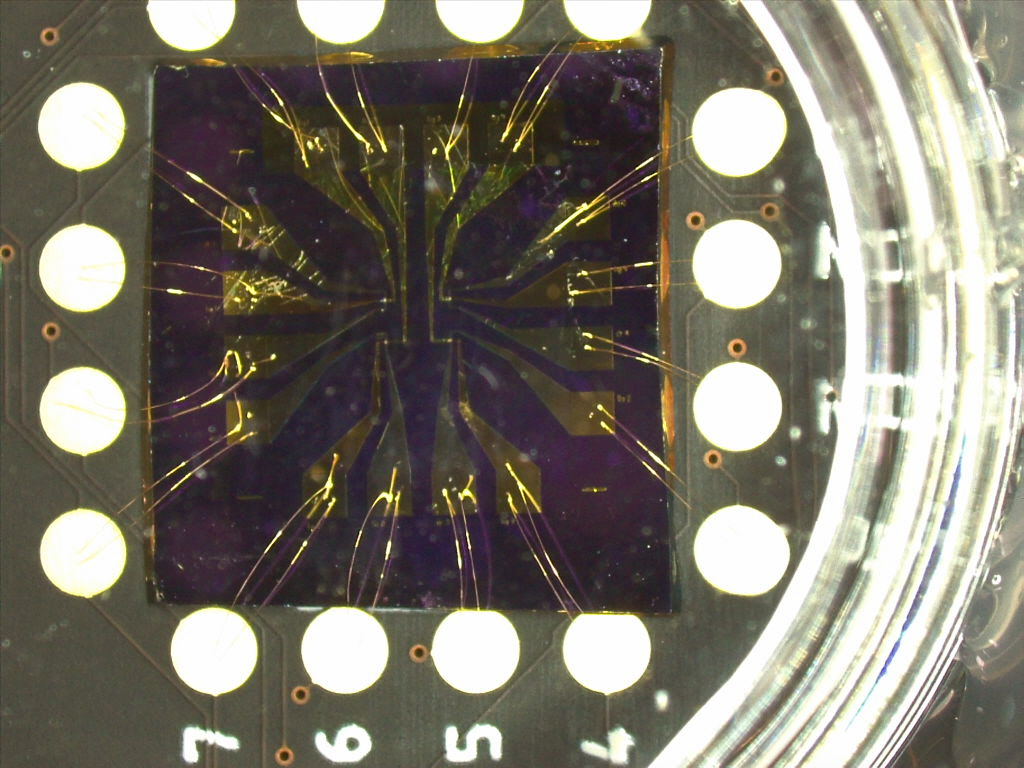}}\\[-0.11in]
	\subfigure[]{\includegraphics[width=0.225\textwidth]{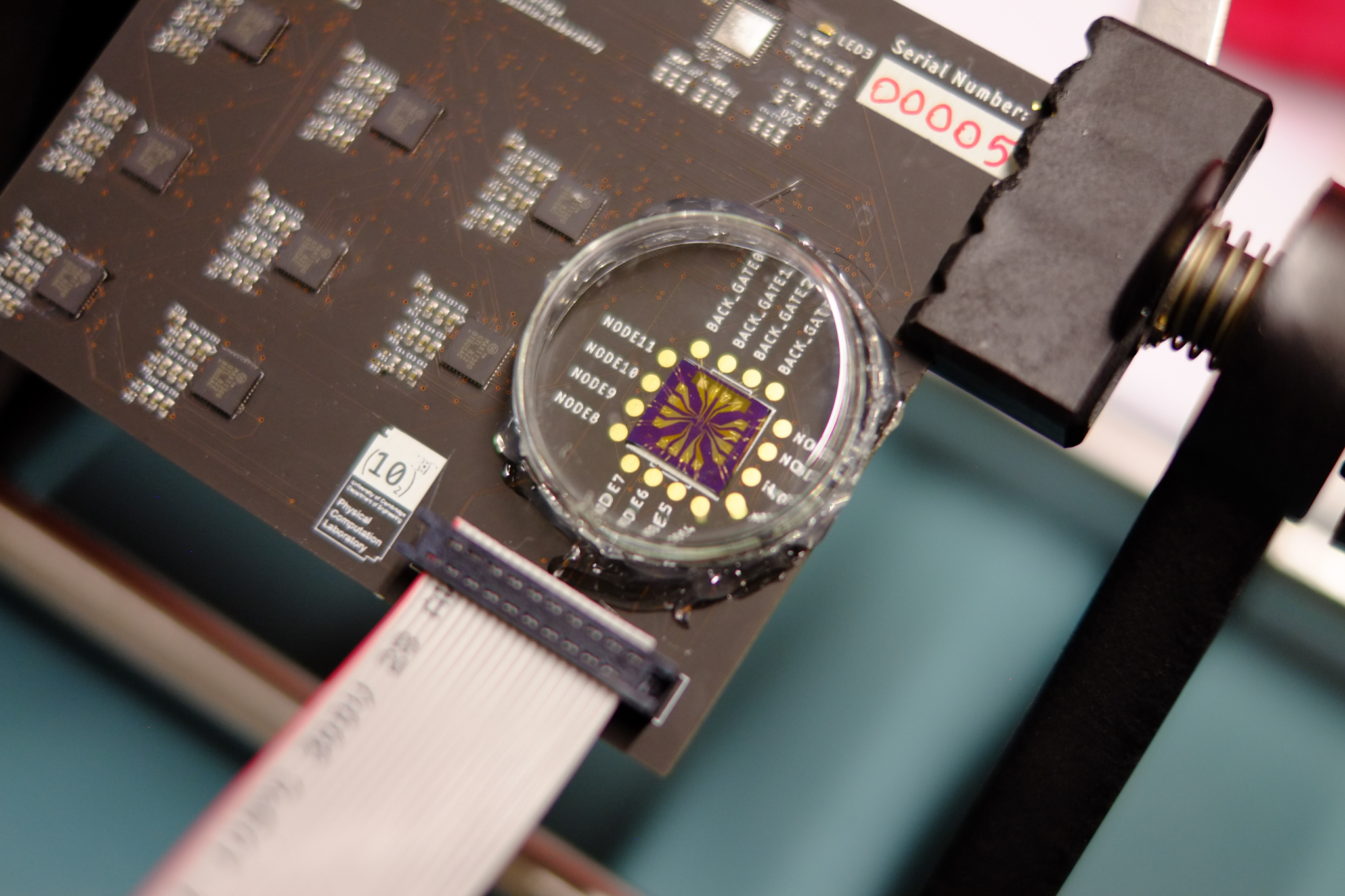}}
	\subfigure[]{\includegraphics[width=0.225\textwidth]{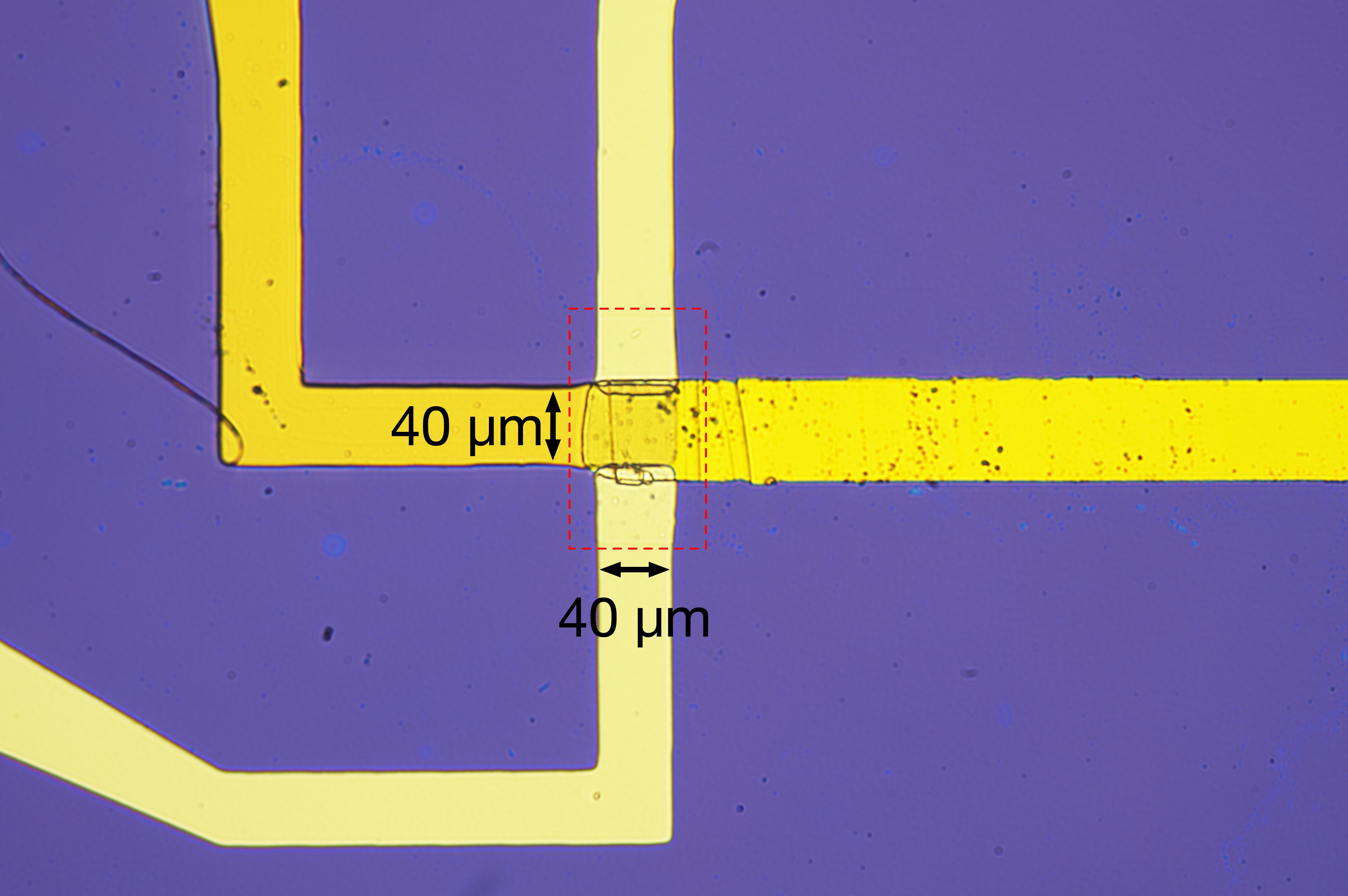}}
	\vspace{-0.1in}
	\caption{\textbf{(a)} Three GFET array dice, each comprising
	four GFETS each having source, drain, top gate and back-gate
	contacts; \textbf{(b)} complete GFET die used in this
	investigation; \textbf{(c)} the custom PCB for dynamic
	reconfiguration of GFET circuits; \textbf{(d)} microscope
	image of the GFET investigated in this paper.}
	\vspace{-0.25in}
	\label{figure:GFETPhoto}
\end{figure}

We fabricate four identical GFETs on each silicon die
(Figure~\ref{figure:GFETPhoto}(a)) and we electrically connect the GFETs
via wire-bonding the die from its gold contacts (Figure~\ref{figure:GFETPhoto}(b))
to a custom printed circuit board (PCB) (Figure~\ref{figure:GFETPhoto}(c)).
The PCB comprises an array of DACs, ADCs, and analog switches,
all of which allow for dynamic and in-situ (re)configuration of a
given circuit. Because graphene is sensitive to atmospheric
effects\footnote{In principle, atmospheric effects can lead to
doping of the channel. These effects should however not occur even
in the absence of the sealed glass protective cover, as we
fabricated the devices in a cleanroom environment and encased the
graphene in alumina as described above. We however cannot rule out
inadvertent doping as a result of the fabrication process.},
and also to protect the wire-bonding, we place a glass protective cover over
each die once bonded to the PCB, sealed with hot glue. The hardware
prototype in Figure~\ref{figure:GFETPhoto}(c) implements the GFETs
required to realise the circuit in Figure~\ref{figure:CircuitEg}, as well as
the analog switching matrix described in Figure~\ref{figure:MAX11}.

\begin{figure}[t]
	\centering
	\subfigure[]{\includegraphics[trim=2cm 5.1cm 2cm 5.5cm, clip=true, width=0.49\linewidth]{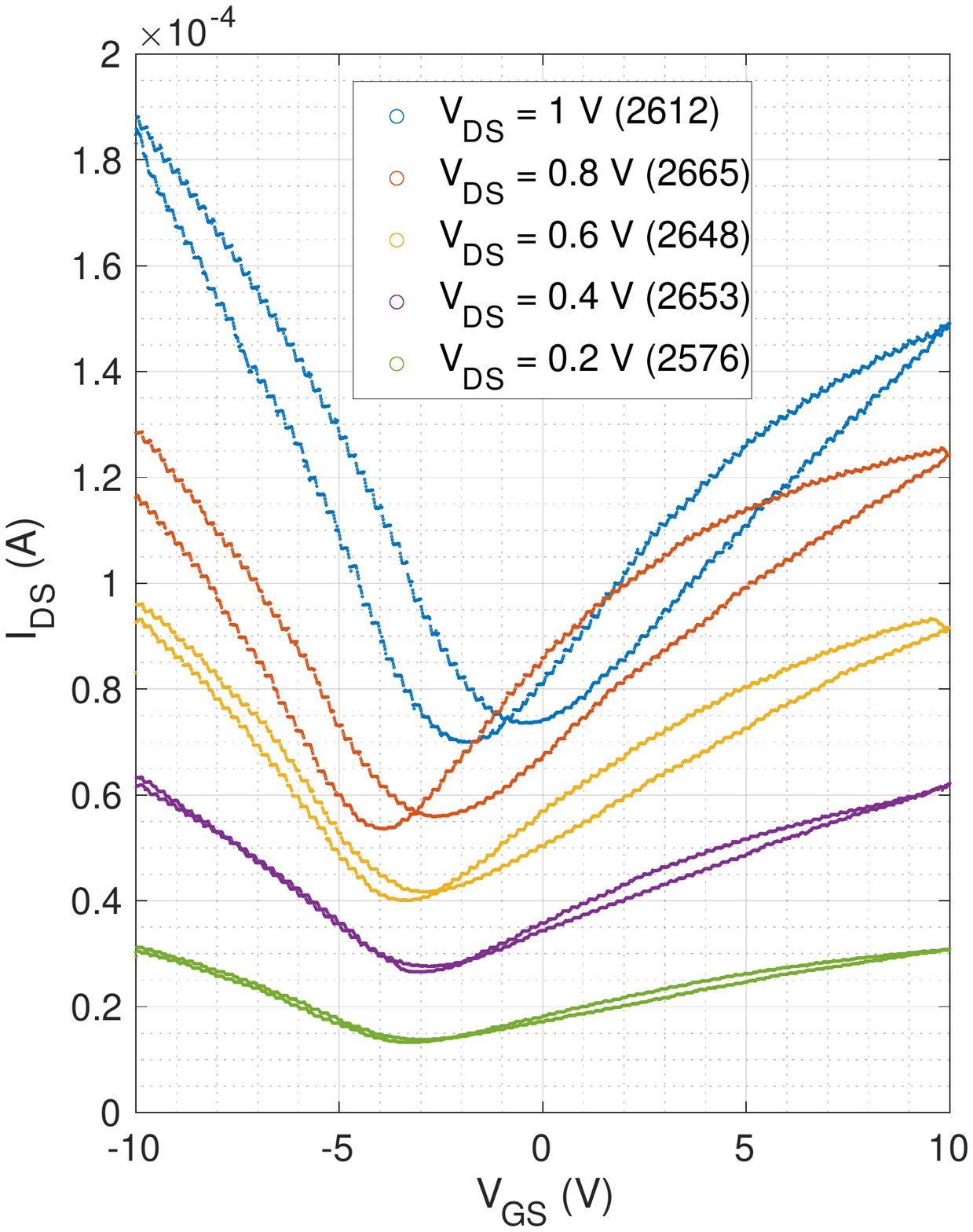}}\llap{\makebox[0.75cm][l]{\raisebox{0.5cm}{\includegraphics[trim=0.1cm 0cm 0.5cm 0cm, clip=true, width=0.75cm]{Illustrations/Grappa00004_GFET3_20x_Annotated.jpg}}}}
	\subfigure[]{\includegraphics[trim=2cm 5.1cm 2cm 5.5cm, clip=true, width=0.49\linewidth]{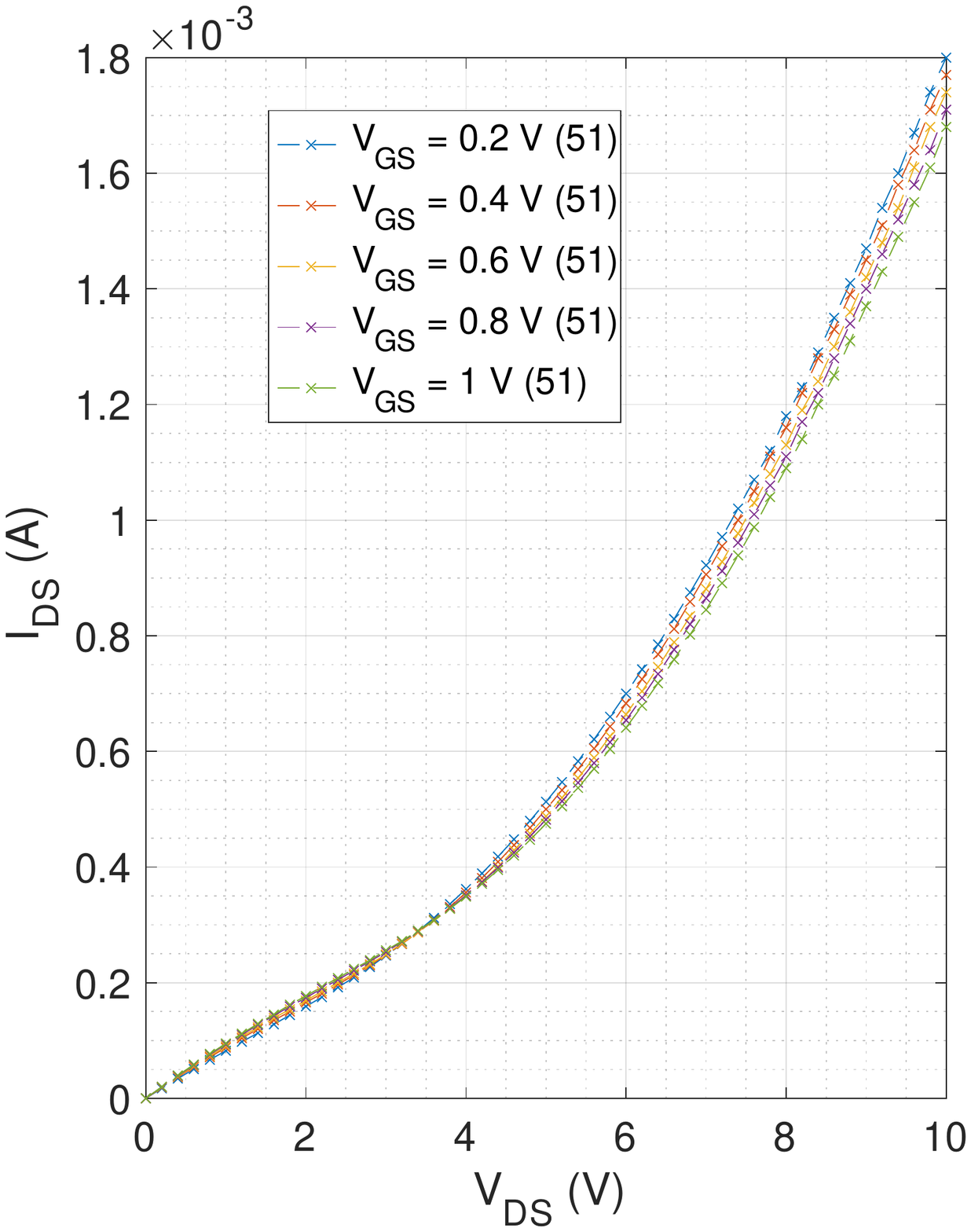}}\llap{\makebox[0.5cm][r]{\raisebox{0.5cm}{\includegraphics[trim=0.1cm 0cm 0.5cm 0cm, clip=true, width=0.75cm]{Illustrations/Grappa00004_GFET3_20x_Annotated.jpg}}}}
	\vspace{-0.2in}
	\caption{\textbf{(a)} Scatter plot of the drain current, $I_{DS}$ against the
	top gate voltage, $V_{GS}$, for different bias voltages of
	$V_{DS}$;  \textbf{(b)} plot of the drain current, $I_{DS}$ against
	the source-drain voltage, $V_{DS}$ for a stepped gate-source
	voltage $V_{GS}$. The number of datapoints for each measurement is in brackets in the legend.}
	\vspace{-0.27in}
	\label{figure:GFETElectrical}
\end{figure}

We performed electrical characterisation of the GFETs using two
Keithley 2450 source-measure units (SMUs) synchronised using TSP-link.
Figure~\ref{figure:GFETElectrical}(a) shows the transfer characteristic
characterisation results of fabricated GFETs, with data
obtained by conducting a linear sweep of the (top) gate-source
($V_{GS}$) voltage between $-$10.0\,V and $+$10.0\,V in both forward and
reverse directions and measuring the resultant drain current
($I_{DS}$). Each curve shows the transfer characteristic for a
constant source-drain bias voltage ($V_{DS}$), which we updated for
each measurement.

The Dirac points in Figure~\ref{figure:GFETElectrical}(a) lie to
the left of 0.0\,V, which suggests an n-type doping of the graphene
channel.  The deepening of the valley with increasing bias voltage
$V_{DS}$ is a commonly-observed characteristic in
GFETs~\cite{Kedzierski2008}, as is the hysteresis in the transfer
characteristics, which the measurements of Figure~\ref{figure:GFETElectrical}
(a) show for all applied bias voltages.
This hysteresis is a result of multiple phenomena: charge trapping
between the graphene channel and the insulating layers are a major cause~\cite{Lemme2010},
however, additional factors include capacitive gating causing a
negative shift and charge transfer causing a positive shift~\cite{Wang2010}
in the conductance with respect to gate voltage.

We also measured the $I_{DS}$ versus $V_{DS}$ characteristics of the
GFETs while varying the gate voltage $V_{GS}$ between 0.2\,V and
1.0\,V, for source-drain voltages $V_{DS}$ over the range 0.0\,V
to 10.0\,V. Traditional MOSFETs exhibit saturation of their $I_{DS}$
versus $V_{DS}$ characteristics, with the characteristics separated
into two main operating regions: linear and nonlinear. In GFETs
however, this saturation does not appear, due to a combination of
graphene's lack of a band gap and Klein tunnelling~\cite{Meric2008}.
Figure~\ref{figure:GFETElectrical}(b) shows the measured characteristics 
for the GFET.

The characterisation data in Figure~\ref{figure:GFETElectrical}(b)
indicate that the devices switch from a relatively linear region
of conductance to a nonlinear region at a bias voltage of around
3.5\,V. This is in line with previous results which show that GFETs,
in comparison to MOSFETs, often have a second linear region; there
is a point of inflection~\cite{Schwierz2010} which appears to be
the case in the plots here, at a V$_{DS}$ of approximately 3.5\,V.

As we show in Section~\ref{section:gfet-simulation-results}, the
nonlinearity of the GFET transfer characteristics, combined with
the tunability of the characteristic shape by controlling $V_{DS}$
allows us to use one or more GFETs to transform uniform
distributions of $V_{GS}$ into non-uniform distributions of $I_{DS}$.

%% file: gfet-simulation-results.tex
\begin{figure}[t]
	\centering
	\subfigure[]{\includegraphics[trim=2.5cm 7.5cm 3cm 8cm, clip=true, width=0.49\linewidth]{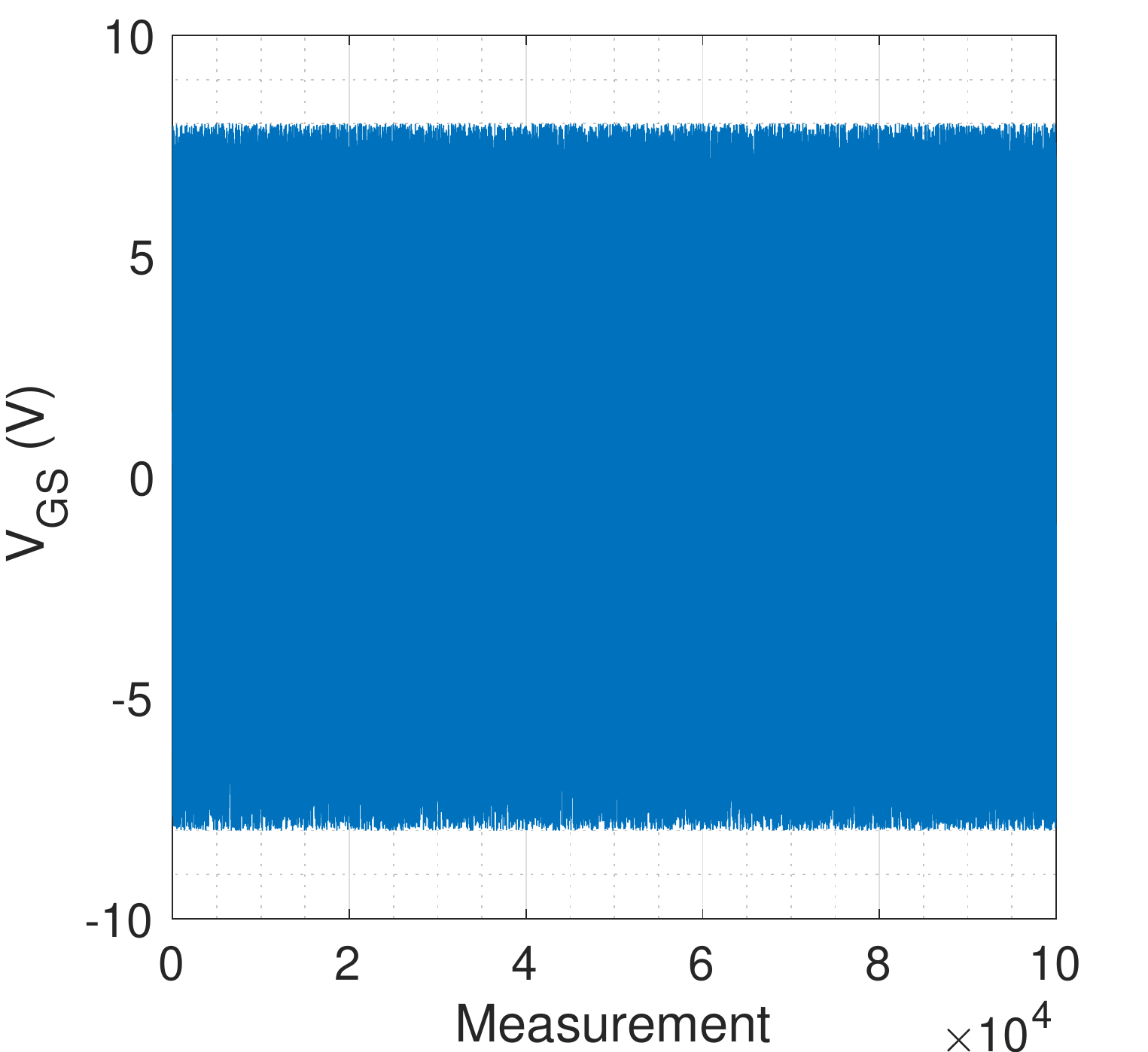}}
	\subfigure[]{\includegraphics[trim=2.5cm 7.5cm 3cm 8cm, clip=true, width=0.49\linewidth]{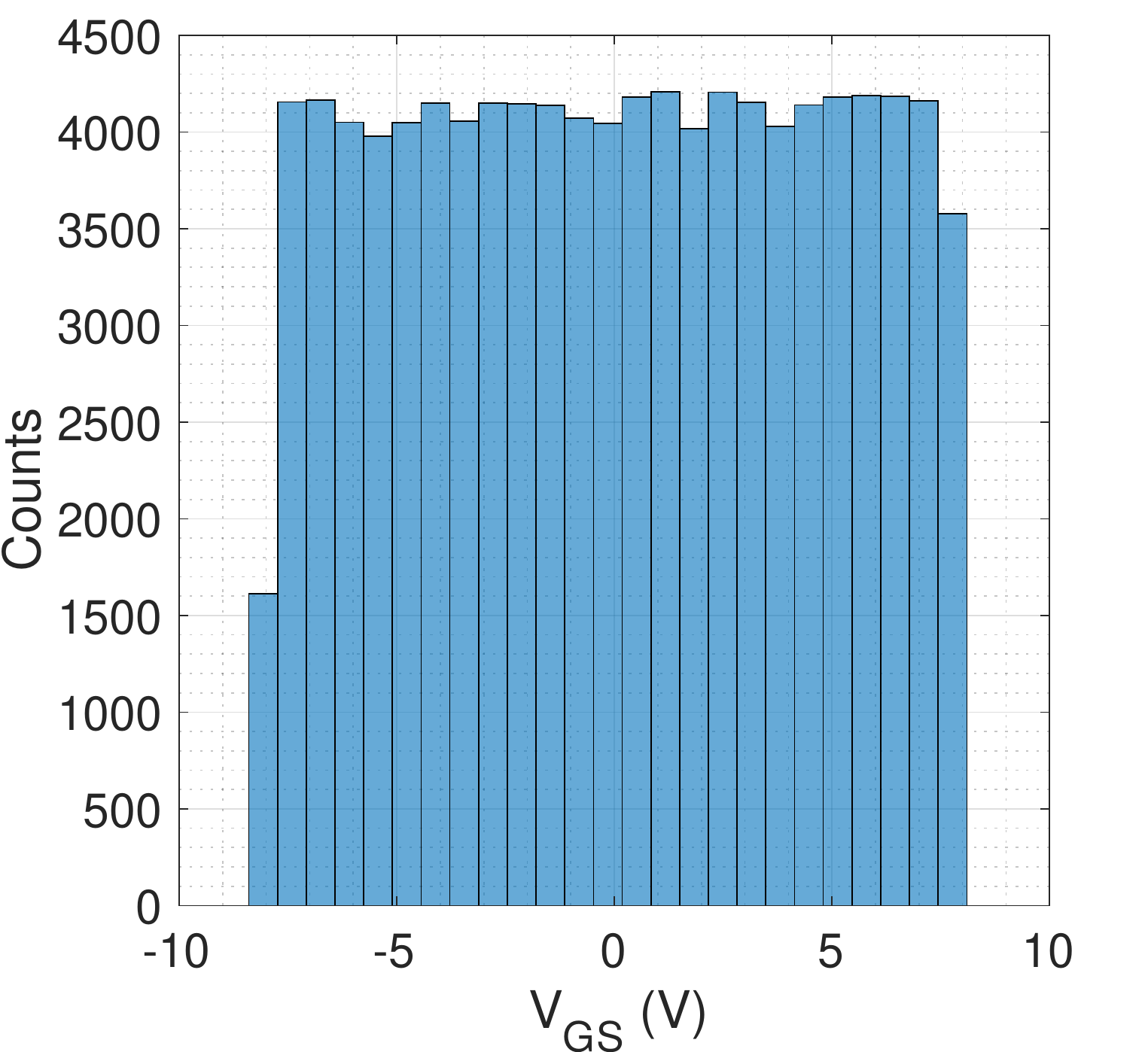}}\\[-0.14in]
	\subfigure[]{\includegraphics[trim=3cm 7.6cm 3cm 8cm, clip=true, width=0.49\linewidth]{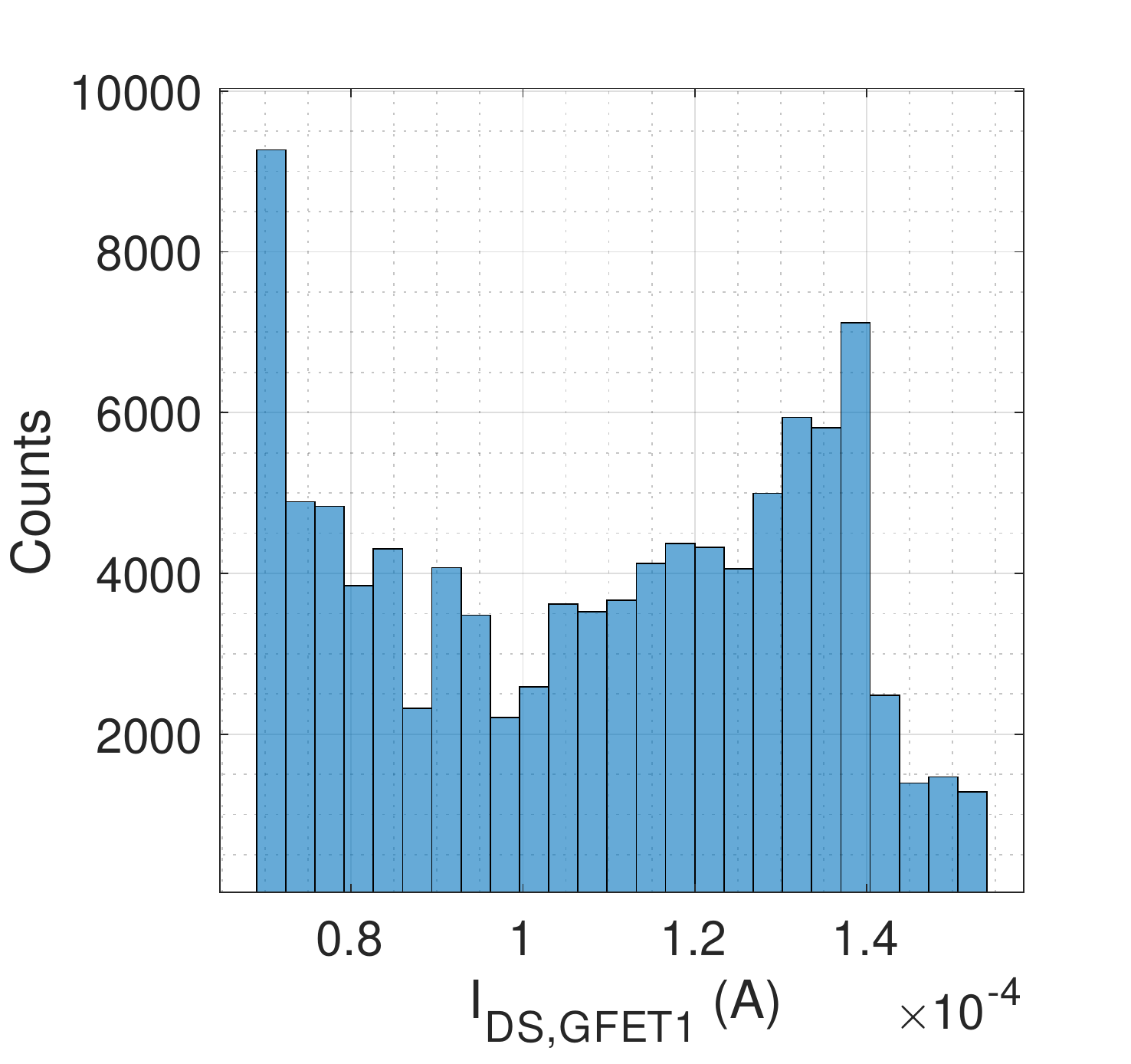}}
	\subfigure[]{\includegraphics[trim=3cm 7.6cm 3cm 8cm, clip=true, width=0.49\linewidth]{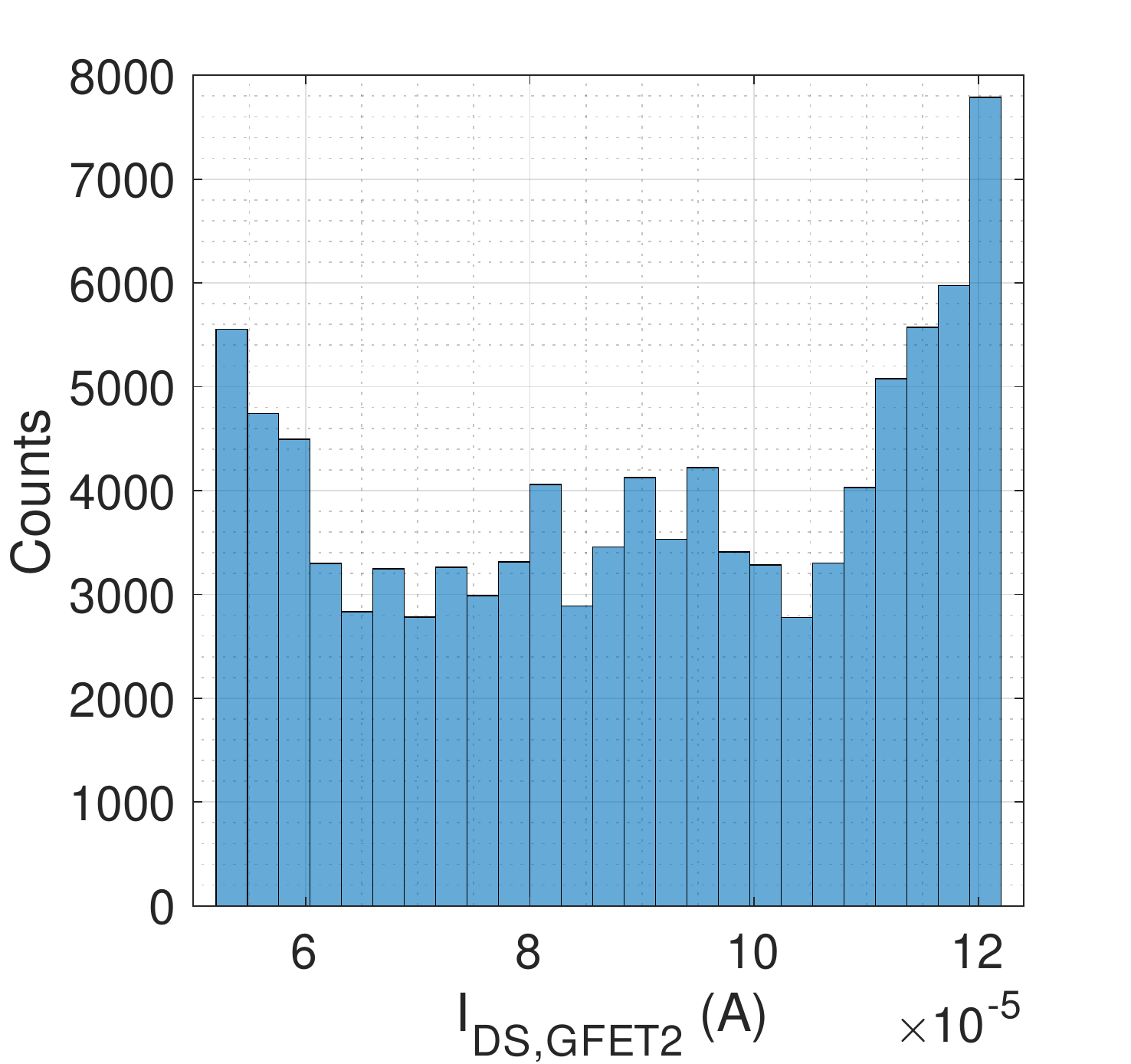}}
	\vspace{-0.2in}
	\caption{\textbf{(a)} Example of a generated uniform pseudo-random 
		voltage distribution used the simulation; \textbf{(b)} histogram of 
		the voltage distribution. \textbf{(c)} and \textbf{(d)}  show histograms of the GFET current distributions for biases of 1\,V and 0.8\,V respectively.}
	\vspace{-0.26in}
	\label{figure:UniformDist}
\end{figure}

\section{Simulations of GFET Circuits}
\label{section:gfet-simulation-results}
\vspace{-0.05in}
We use the GFET characterisation data from
Section~\ref{section:GFET-fabrication} to simulate possible topologies
for the GFET-based PRVA of
Figure~\ref{figure:CircuitEg}, using a custom-built simulation model
of the circuit, implemented in Mathematica. We use an interpolating
function whose values are determined from the measured GFET transfer characteristics 
and stimulate the gate of the first GFET in the circuit with a
uniform random distribution between $-$8 and $+$8, effectively a $V_{GS}$
voltage in the range $-$8\,V to $+$8\,V. Figure~\ref{figure:UniformDist}
shows the time series of the uniform random voltages and their
corresponding histogram distribution. We pass the output of the
model of the first GFET, which is its drain current, through a
modelled resistor which converts the drain current into a voltage.
We then apply this voltage to the model of the second GFET of
Figure~\ref{figure:CircuitEg}, which we again model by encapsulating
our experimental measurement data in another interpolating function.
We apply the drain current of the second GFET through another resistor
to obtain the output voltage.

\begin{figure}[t]
	\centering
	\subfigure[]{\includegraphics[trim=3cm 7.6cm 3cm 7.5cm, clip=true, 
	width=0.49\linewidth]{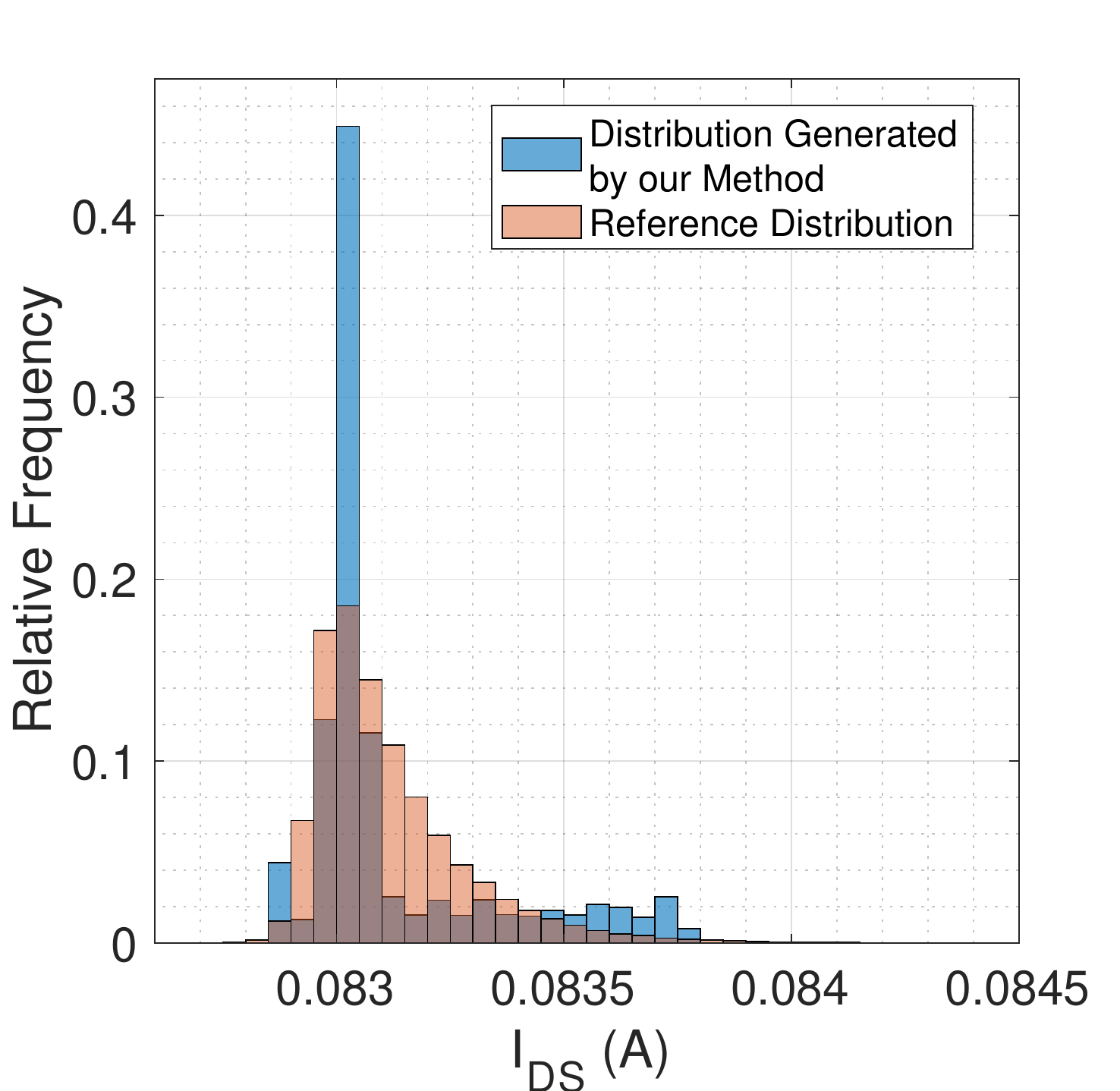}}
	\subfigure[]{\includegraphics[trim=2.5cm 7.6cm 3cm 7.5cm, clip=true, 
		width=0.49\linewidth]{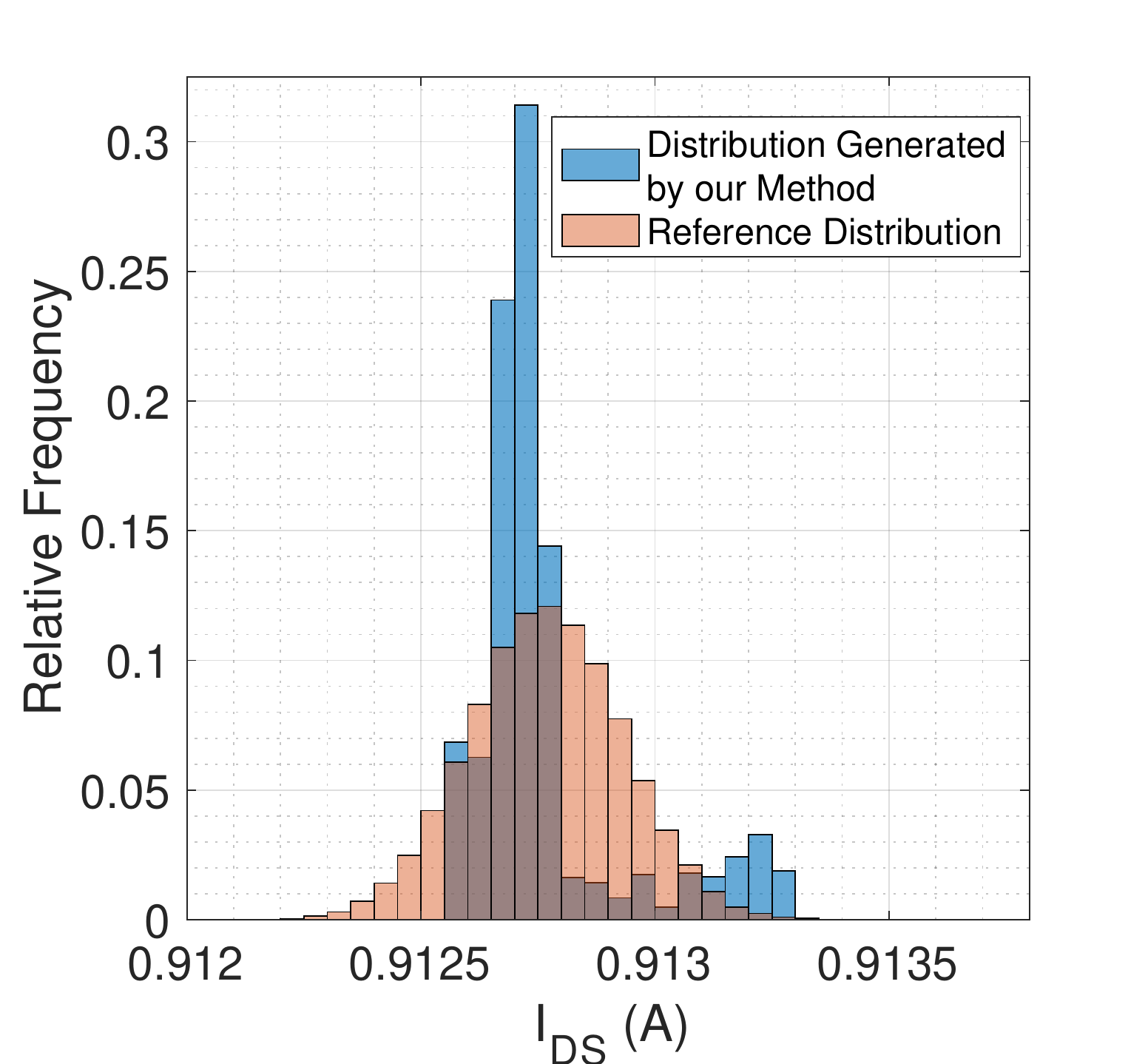}}
	\vspace{-0.2in}
	\caption{Histograms of simulated transformations of uniform to 
		non-uniform distributions. \textbf{(a)} GFET-only based transformation 
		with a reference Burr Type-XII; \textbf{(b)} Combined GFET and 
		computational transform with reference lognormal.}
	\vspace{-0.28in}
	\label{figure:CircuitOut}
\end{figure}

We used the characteristics of the GFET in
Figure~\ref{figure:GFETElectrical} for simulations here. For the first GFET,
we used the characteristic for a 0.8\,V bias and for the second GFET, we 
used the characteristic for a 1\,V bias. We use a resistance of
2.2k\,$\Omega$ for the first resistor and of 1k\,$\Omega$ for the 
output resistance. Figure \ref{figure:UniformDist} (c) shows the distribution
of currents from the first GFET, and Figure \ref{figure:UniformDist}
(d) shows the distribution for the second GFET.
Figure~\ref{figure:CircuitOut} (a) shows final output distribution
(i.e., the output of the first GFET passed through the second GFET). 

We investigated the possibility of combining the GFET-based distribution
with additional subsequent software transformation by running a second 
simulation of the circuit in Figure~\ref{figure:CircuitEg} using the 
experimentally measured GFET data, with the characteristic for a 
1\,V biased GFET as the first GFET and the characteristic for a 
0.8\,V biased GFET as the second GFET. We chose a resistance of 
1.2k\,$\Omega$ for the first resistor and 1k\,$\Omega$ for the 
output resistance. The initial output distribution was skewed to 
the right, and so we took the complementary cumulative distribution function, 
$\bar{F}(x)$, given by:

\begin{equation}
	\bar{F}(x) = 1 - F(x),
\end{equation}
where F(x) is the output distribution of the circuit. To compare 
the similarity of the generated distribution to a genuine lognormal, we calculated 
the mean ($\mu$) and standard deviation ($\sigma$) of the logarithm 
of the simulation output and used these as parameters to 
generate a lognormal distribution over the same input space. 
Figure~\ref{figure:CircuitOut} (b) shows histograms of the simulation 
output and the reference distribution.

%% file: monte-carlo-results.tex
\section{End-to-End Example: Monte Carlo Integration}
\label{section:Monte-Carlo-results}
\vspace{-0.05in}
It is common to integrate un-normalised non-uniform density functions to convert them to valid probability density functions~\cite{chen2017monte}.
Monte Carlo integration is a convenient way of doing this.
Consider a thought experiment where we have collected a one-dimensional dataset. The data describes the probability that indoor
air quality is below acceptable levels given the reading from a volatile organic compound gas sensor.
This dataset is well described by a lognormal distribution.
We calculate the mean $\mu$ and standard deviation $\sigma$ of the data.
Before we can interrogate the data to ask questions such as
``What is the probability that the air quality is unacceptable given that the sensor reading is 25?'',
we must find a normalisation constant to ensure that the probability density function integrates to one.
The normalisation requires the integration of a lognormal distribution
$f(x)$ where $Z$ is an unknown normalising constant, $\mu = 0$ is the mean
and $\sigma = 0.25$ is the standard deviation:

\begin{align}
f(x) = \frac{Z}{x} e^{-\frac{(\ln (x) -\mu)^2}{2 \sigma^2}}.
\end{align}

Let $E$ be the error of the Monte Carlo integration and $t$ be the
time taken by the integration. Let $N$ be the number of random
samples used in the integration and $D$ be the distribution that
we sample from. Let $A$ be the area of each rectangle
used in the integration and $b$ and $h$ be the corresponding rectangle
base and height. Algorithm~\ref{Algorithm3} shows the integration
scheme.

We repeated the integration with $D$ as: 1) a hardware-generated lognormal
distribution and 2) a lognormal distribution generated with the C++ standard
library utility for generating lognormal variates, with the same $\mu$ and
$\sigma$ as $f$. We also performed the integration with $D$ as a uniform
distribution generated with the C++ standard library random number
generator, with various ranges. We assume that samples from the hardware
lognormal generator can be generated in the time required for one memory
access. We ran all simulations on a 2.8 GHz Intel Core i7 CPU using OpenMP
parallelisation to utilise all eight processor threads.

\begin{algorithm}[h]
\DontPrintSemicolon
\SetAlgoLined
\KwResult{Error $E$ and time $t$}
Timer start\;
Generate $N$ random samples from distribution $D$\;
Sort $N$ random samples\;
\For{$i = 1$ to $N$}{
$b$ = Samples[$i$] $-$ Samples[$i-1$]\;
$h$ = (f(Samples[$i$]) $+$ f(Samples[$i-1$]))/2\;
$A += b*h$\;
}
$E$ = |$1-A$|\;
Timer stop\;
t = stop $-$ start\;
RETURN $E,t$\;
\caption{Monte Carlo integration. Array index begins at zero.}
\label{Algorithm3}
\end{algorithm}
\vspace{-0.27in}
\subsection{Results}
\begin{figure}[]
	\centering
	\subfigure[]{\includegraphics[trim= 0cm 0.1cm 0cm 0cm, clip=true,  width=0.480\linewidth]{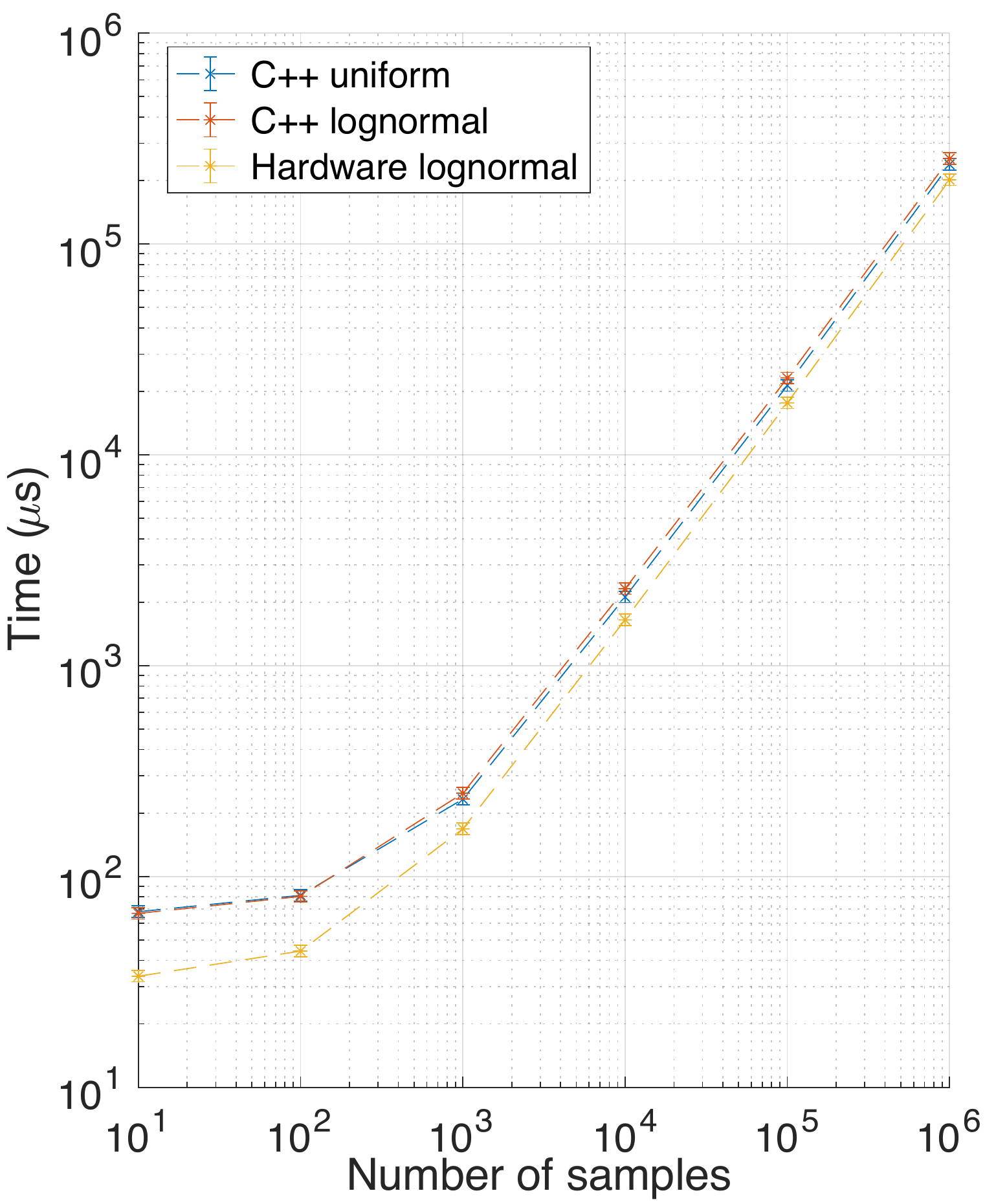}}
	\subfigure[]{\includegraphics[trim= 0cm 0.1cm 0cm 0cm, clip=true,  width=0.480\linewidth]{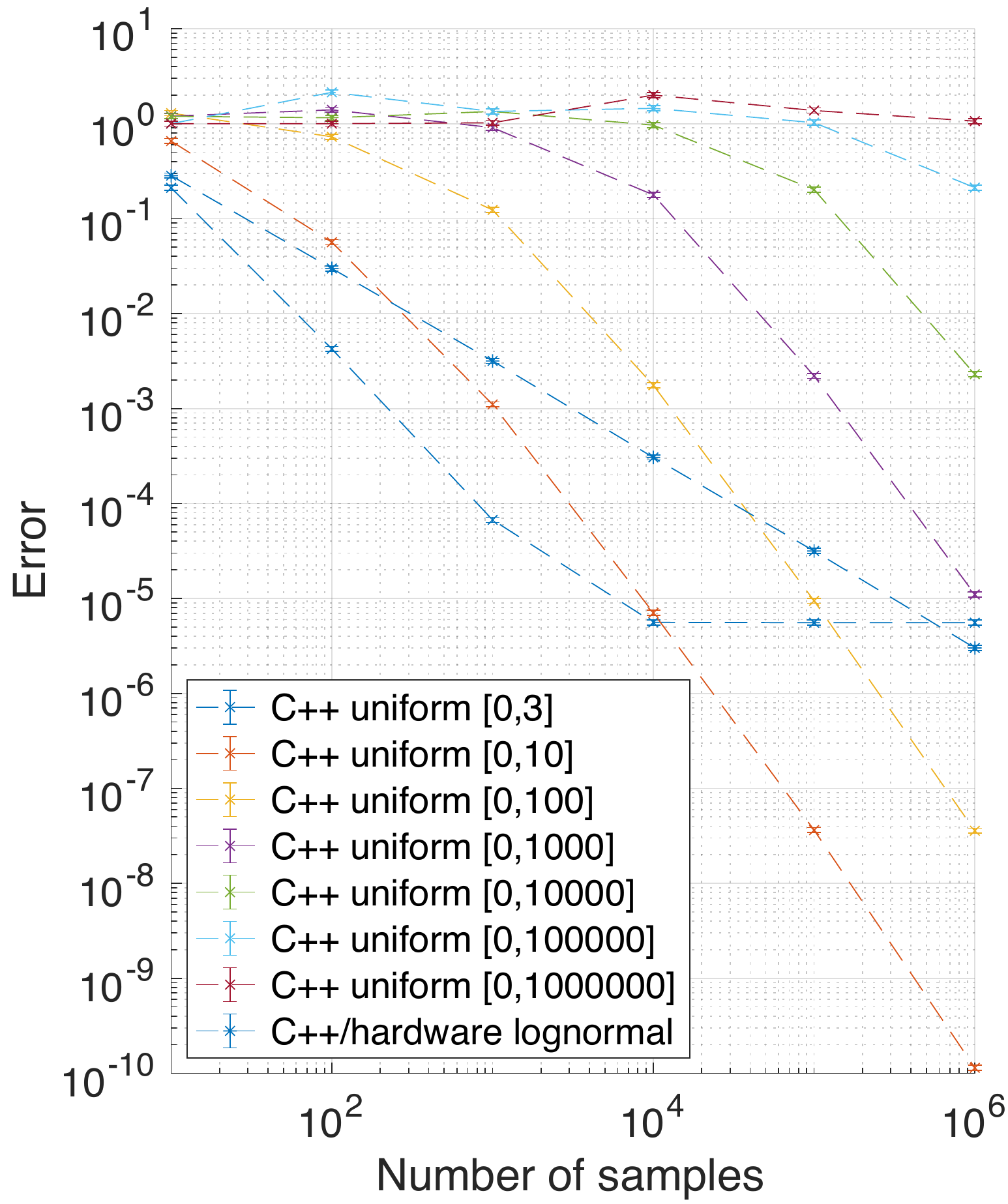}}
	\vspace{-0.11in}
	\caption{\textbf{(a)} Time taken to perform a Monte Carlo
	simulation for $N$ samples using the C++ uniform and lognormal
      random number generators and the proposed hardware
	random number generator.; \textbf{(b)} Error in the numerical
	integration produced by a Monte Carlo simulation for $N$
	samples using the C++ uniform and lognormal random number
	generators. The error bars show a 90 \% confidence interval on the mean
	of 1000 samples for each point.}
	\vspace{-0.21in}
	\label{figure:MonteCarloResults}
\end{figure}

\vspace{-0.05in}
Figure~\ref{figure:MonteCarloResults}(a) shows that it is on
average $1.05 \times$ faster to use a C++ uniform random number
generator than a C++ lognormal random number generator.
Running the program assuming that the lognormal
samples are generated by the hardware random number generator in the same
amount of time required for a memory access is up to
$1.99 \times$ faster and always at least $1.26 \times$ faster than using the
C++ lognormal random number generator.
Figure~\ref{figure:MonteCarloResults}(b) shows that the error reduction
for the Monte Carlo integration using the [0, 3] C++ uniform random
number generator plateaus at around $10^4$ samples but for the
hardware lognormal the error continues to decrease. The lognormal
and [0, 3] C++ uniform lines intersect at between $10^5$ and $10^6$
samples and the intersection point shifts to the right as we increase
the range of the C++ uniform random number generator.

\subsection{Insights from Monte Carlo Integration}
Figure~\ref{figure:MonteCarloResults}(b) shows that increasing the range of the
C++ uniform random number generator decreases the minimum error that the integration
plateaus at. Unfortunately, increasing the range of the C++ uniform random number generator
also increases the number of samples required for the estimate of the area to
approach the true value. The proportion of the uniform probability
density function overlapping the lognormal probability density function increases
as the range of the uniform distribution is increased. For example, the overlap of the [0, 3]
uniform distribution and $f(x)$ is smaller than that for the [0, 10] uniform distribution and $f(x)$.
Therefore, the error can never be zero with any uniform distribution other than [0, $\infty$]
as the source of samples for the numerical integration of $f(x)$.
For a given function we cannot know beforehand which range of uniform random numbers will produce a
sufficiently small bound on the error of integration.
The sharp cut-off of the uniform distribution will always cause us to miss some of the area under $f(x)$ that we want to integrate.
We can avoid this problem by sampling from the smooth lognormal density that we want to integrate.
When performing the Monte Carlo integration of any non-uniform distribution
we should sample from the probability density function of that distribution with
the same parameters to minimise the error and number of samples required to get
a reasonable estimation. This is not possible when sampling from a bounded
uniform distribution.

\subsection{The Accept-Reject Method and Computational Cost}

The computational complexity of the operations we perform in Algorithm~\ref{Algorithm3} with the exception of the sort have computational cost $\mathcal{O}(N)$
where $N$ is the number of samples.
The computational cost of sorting is $\mathcal{O}(N\log(N))$.
We expect larger performance gains for integration schemes that do not require a sort operation. 
Future work will use area-based ``dart throwing'' integration schemes that avoid sorting.

We have shown that it is better to integrate $f(x)$ with samples from a non-uniform distribution with a similar shape.
The accept-reject method is the most flexible software method for generating samples from an arbitrary non-uniform distribution.
The method works by transforming samples from a uniform distribution (that the computer can directly generate samples from) 
to a target non-uniform distribution.
We choose to accept or reject samples based on a probabilistic comparison between a uniform random sample 
and the ratio of the target and uniform probability density functions.
As this comparison is probabilistic, it is possible to become stuck rejecting samples forever~\cite{devroye2008non}.

%% file: relatedwork.tex
\section{Related Research}
\label{Section:RelatedRedearch}
\vspace{-0.05in}

A hardware random number generator, integrated in a CPU capable of producing samples from arbitrary distributions does not currently exist.
Table~\ref{Table:StateOfTheArt} shows the state-of-the-art of hardware non-uniform random number generators.
The prior work on non-uniform random number generation in Table~\ref{Table:StateOfTheArt} is fundamentally different to prior work on uniform random number generation.
The publications in Table~\ref{Table:StateOfTheArt} characterise the non-uniform distribution of the physical process used to obtain the random samples.
The prior work on uniform random number generators does not produce or refer to a non-uniform distribution of random numbers~\cite{wallace2016toward, lee2019true, perach2019asynchronous, srinivasan20102}.
The uniform random number generation efforts excluded from Table~\ref{Table:StateOfTheArt} produce single-bit samples where the result is either 0 or 1.
The non-uniform random number generation efforts included in Table~\ref{Table:StateOfTheArt} produce multiple (usually 6 or greater) bit samples with a given non-uniform distribution.
There is only one prior work from this decade that is capable of producing samples from arbitrary non-uniform distributions~\cite{nguyen2018programmable}.
Their method is not well-suited for integration with current CPU architectures as it requires large optical components.
In contrast, Section~\ref{section:GFET-prva} describes how the GFET random number generator would interface with an existing CPU.

Beyond its application to Monte Carlo methods, the GFET-based
non-uniform random number generator could also see use in other
compute-intensive analyses~\cite{10.1145/1989493.1989506}, as well
as in techniques that trade fidelity of their results for
lower power dissipation~\cite{2016:CSP:2901318.2901347}.

Our approach and the approach by Nguyen et al.~\cite{nguyen2018programmable}
differ from the state-of-the-art in the following important way:
They are programmable to allow an arbitrary distribution to be achieved.
In contrast, most of the designs in Table~\ref{Table:StateOfTheArt} are not programmable~\cite{jiang2017novel, marangon2017source, avesani2018source, raffaelli2018homodyne, tomasi2018model, park2019practical, lin2019true, guo2019parallel, xu2019spad}.
These are only capable of producing samples from one type of distribution and can only do this for one set of fixed parameters, so would require external modification to make them programmable.

In contrast, Zhang et al.~\cite{zhang2018architecting} can produce an exponential distribution with any parameters, but would require external modification to produce samples from distributions with different shapes.
The work by Hu et al.~\cite{hu2019gaussian} and Meech et al.~\cite{meech2020efficient} can produce samples from a normal distribution with any parameters but would require external modification to allow them to produce samples from distributions with different shapes.
The non-programmable random number generators in Table~\ref{Table:StateOfTheArt} were designed to suit applications such as cryptography which are drastically different to the applications we propose for our architecture.

Currently no architecture exists with a Gb/s generation rate for arbitrary
non-uniform distributions. As the method we present is analog,
it should be possible to use existing technology to drive
and sample from it.  This will allow us to produce samples from
arbitrary distributions at Gb/s sample rates.
Popular statistical tests such as Dieharder are designed for
samples from a [0, 1] uniform distribution and are therefore not compatible with
the non-uniformly distributed random numbers produced in this work~\cite{Dieharder}.

\begin{table}[t]
\caption{State-of-the-art in uniform and non-uniform random number
generation architectures. In contrast to the methods below,
the method we present in this paper generates arbitrary distributions
and is only limited by the speed of available analog-to-digital
converters (ADCs).}
\vspace{-0.1in}
\centering
\begin{tabular}{l|llcl}
\toprule
\rowcolor{a}\bf{Architecture}		& \bf{Speed}	& \bf{Distribution(s)} & \bf{Year} & \bf{Paper}					\\
\midrule
\rowcolor{b}Memristor				 & 6.00\,kb/s		& Unnamed		    & 2017 & \cite{jiang2017novel}			      \\
\rowcolor{a}Photon Detection & 1.77\,Gb/s		& Exponential		& 2017 & \cite{marangon2017source}		  	\\
\rowcolor{b}FRET				     & 2.89\,Gb/s		& Exponential		& 2018 & \cite{zhang2018architecting}			\\
\rowcolor{a}Photo Diode			 & 17.4\,Gb/s		& Husumi		    & 2018 & \cite{avesani2018source}		     	\\
\rowcolor{b}Photon Detection & 66.0\,Mb/s		& Arbitrary		  & 2018 & \cite{nguyen2018programmable}		\\
\rowcolor{a}Photon Detection & 200\,Mb/s		& Normal    	  & 2018 & \cite{raffaelli2018homodyne}	  	\\
\rowcolor{b}Photon Detection & 320\,Mb/s		& Exponential		& 2018 & \cite{tomasi2018model}	  	      \\
\rowcolor{a}Electronic Noise & 6.40\,Gb/s		& Normal    		& 2019 & \cite{hu2019gaussian}	  	      \\
\rowcolor{b}Photon Detection & 6.80\,Mb/s		& Exponential		& 2019 & \cite{park2019practical}	  	    \\
\rowcolor{a}Photon Detection & 63.5\,Mb/s		& Exponential		& 2019 & \cite{lin2019true}	  	          \\
\rowcolor{b}Photon Detection & 8.25\,Mb/s		& Normal    		& 2019 & \cite{guo2019parallel}	  	      \\
\rowcolor{a}Photon Detection & 1.00\,Mb/s		& Exponential		& 2019 & \cite{xu2019spad}	  	          \\
\rowcolor{b}Electronic Noise & 13.8\,kb/s		& Normal    		& 2020 & \cite{meech2020efficient}	  	  \\

\bottomrule
\end{tabular}
\vspace{-0.24in}
\label{Table:StateOfTheArt}
\end{table}

\subsection{Alternative Devices}

Reconfigurable FETs (RFETs) based on nanowires also possess tunable
characteristics: printed multi- and parallel nanowire Schottky barrier
devices have similar transfer characteristics to GFETs, where the majority carriers are
determined by the gate voltage. FinFETs may also be of interest, as their
multi-gate structure allows for tunable characteristics, although not in the same
manner as with GFETs or RFETs. FETs based on carbon nanotubes (CNTs) also 
may have properties of interest, however these also have a number of well-documented 
issues with scalability and uniformity which limits their prospects of commercialisation.

FinFETs on a 14\,nm CMOS process were used by Matthew et al. for a hardware RNG~\cite{Matthew2016},
with a throughput on the order of a few hundred Mb/s, however it can only produce samples from a Gaussian distribution,
so it is not programmable in the way that ours is. A CNT-based RNG was demonstrated by Rojas et al. in 2017~\cite{Rojas2017},
however no data on the sample rate was produced, and the generated distribution is uniform.
A nanowire-based RNG was demonstrated by He et al.~\cite{He2016}, however this had a generation 
rate of less than 10\,Mb/s and also involved superconducting properties,
requiring operation at temperatures of a few degrees Kelvin and making them impractical for mainstream use.

%% file: conclusions.tex
\section{Summary and Insights}
\label{section:conclusions}
\vspace{-0.05in}
This article demonstrates a novel circuit-level approach to generating
samples from non-uniform probability distributions, exploiting the
transfer characteristics and ambipolarity of graphene field-effect
transistors (GFETs).

We describe the fabrication of arrays of GFETs on a silicon substrate
and wire bonding of the fabricated devices to a custom PCB. 
We experimentally characterise the GFET transfer and output 
characteristics at a range of GFET $V_{DS}$ bias voltage 
configurations. Using the obtained characterisation data, we simulate 
possible circuit designs for PRVAs
comprising circuits requiring just two transistors and a resistor 
(or transimpedance amplifier). The results demonstrate that a circuit 
comprising a chain of two GFETs transforms a 
uniformly distributed random input voltage
into a non-uniformly distributed output. 

We evaluate the end-to-end use of the GFET-circuit-generated
distributions in an application performing Monte Carlo integration.
The results show that, using the GFET-circuit-generated non-uniform
distributions instead of uniform random samples for sampling locations
in the lognormal distribution improves the speed of Monte Carlo
integration by a factor of up to 2\,$\times$. This speedup is based
on the assumption that the analog-to-digital converters that will
be necessary to read outputs from GFET-based random number generation
circuit can produce samples in the same amount of time that it takes
to perform memory accesses.

%% file: ms.bbl
\begin{thebibliography}{10}

\bibitem{avesani2018source}
M.~Avesani, D.~G. Marangon, G.~Vallone, and P.~Villoresi.
\newblock Source-device-independent heterodyne-based quantum random number
  generator at 17 gbps.
\newblock {\em Nature communications}, 9(1):1--7, 2018.

\bibitem{Dieharder}
R.~G. Brown.
\newblock Dieharder.
\newblock Available at:
  \url{http://webhome.phy.duke.edu/~rgb/General/dieharder.php} Accessed
  17/04/2020.

\bibitem{10.1145/1989493.1989506}
V.~Caparr\'{o}s~Cabezas and P.~Stanley-Marbell.
\newblock Parallelism and data movement characterization of contemporary
  application classes.
\newblock SPAA ’11, page 95–104. ACM, 2011.

\bibitem{Chen2008}
J.-H. Chen, C.~Jang, S.~Xiao, M.~Ishigami, and M.~S. Fuhrer.
\newblock Intrinsic and extrinsic performance limits of graphene devices on
  sio2.
\newblock {\em Nature Nanotechnology}, 3(4):206--209, 2008.

\bibitem{Daubechies1992}
I.~Daubechies.
\newblock {\em Ten Lectures on Wavelets}, pages 53--55.
\newblock Society for Industrial and Applied Mathematics, Jan. 1992.
\newblock ISBN: 0898712742.

\bibitem{devroye2008non}
L.~Devroye.
\newblock {Non-Uniform Random Variate Generation}.
\newblock page~42. Springer-Verlag, McGill University Montreal H3A 2K6 Canada,
  1986.
\newblock ISBN: 1461386454.

\bibitem{Rojas2017}
W.~A. Gaviria~Rojas, J.~J. McMorrow, M.~L. Geier, Q.~Tang, C.~H. Kim, T.~J.
  Marks, and M.~C. Hersam.
\newblock Solution-processed carbon nanotube true random number generator.
\newblock {\em Nano Letters}, 17(8):4976--4981, 2017.

\bibitem{Graphenea}
Graphenea.
\newblock Graphenea monolayer graphene film on polymer film.
\newblock
  \url{https://cdn.shopify.com/s/files/1/0191/2296/files/Graphenea_Monolayer_Graphene_on_Polymer_Film_Datasheet_05-25-2020.pdf?v=1591290809},
  2020.

\bibitem{Graps1995}
A.~{Graps}.
\newblock An introduction to wavelets.
\newblock {\em IEEE Computational Science and Engineering}, 2(2):50--61, 1995.

\bibitem{guo2019parallel}
X.~Guo, C.~Cheng, M.~Wu, Q.~Gao, P.~Li, and Y.~Guo.
\newblock Parallel real-time quantum random number generator.
\newblock {\em Optics letters}, 44(22):5566--5569, 2019.

\bibitem{He2016}
Y.~He, W.~Zhang, H.~Zhou, L.~You, C.~Lv, L.~Zhang, X.~Liu, J.~Wu, S.~Chen,
  M.~Ren, Z.~Wang, and X.~Xie.
\newblock Bias-free true random number generation using superconducting
  nanowire single-photon detectors.
\newblock {\em Superconductor Science and Technology}, 29(8):085005, jun 2016.

\bibitem{hu2019gaussian}
Y.~Hu, Y.~Wu, Y.~Chen, G.~C. Wan, and M.~S. Tong.
\newblock Gaussian random number generator: Implemented in fpga for quantum key
  distribution.
\newblock {\em International Journal of Numerical Modelling: Electronic
  Networks, Devices and Fields}, 32(3):e2554, 2019.

\bibitem{IntelHardware}
Intel.
\newblock Intel® digital random number generator (drng), 2018.
\newblock Available at:
  \url{https://software.intel.com/sites/default/files/managed/98/4a/DRNG_Software_Implementation_Guide_2.1.pdf}
  Accessed 17/04/2020.

\bibitem{jiang2017novel}
H.~Jiang, D.~Belkin, S.~E. Savel’ev, S.~Lin, Z.~Wang, Y.~Li, S.~Joshi,
  R.~Midya, C.~Li, M.~Rao, et~al.
\newblock A novel true random number generator based on a stochastic diffusive
  memristor.
\newblock {\em Nature communications}, 8(1):1--9, 2017.

\bibitem{Kedzierski2008}
J.~{Kedzierski}, P.~{Hsu}, P.~{Healey}, P.~W. {Wyatt}, C.~L. {Keast},
  M.~{Sprinkle}, C.~{Berger}, and W.~A. {de Heer}.
\newblock Epitaxial graphene transistors on sic substrates.
\newblock {\em IEEE Transactions on Electron Devices}, 55(8):2078--2085, Aug
  2008.

\bibitem{keyes1985makes}
R.~W. Keyes.
\newblock What makes a good computer device?
\newblock {\em Science}, 230(4722):138--144, 1985.

\bibitem{kullback1951}
S.~Kullback and R.~A. Leibler.
\newblock On information and sufficiency.
\newblock {\em Ann. Math. Statist.}, 22(1):79--86, 03 1951.

\bibitem{lambert2018student}
B.~Lambert.
\newblock {\em A Student’s Guide to Bayesian Statistics}, pages 23--50.
\newblock SAGE, 2018.
\newblock ISBN: 1473916364.

\bibitem{lee2019true}
K.~Lee and M.~Lee.
\newblock True random number generator (trng) utilizing fm radio signals for
  mobile and embedded devices in multi-access edge computing.
\newblock {\em Sensors}, 19(19):4130, 2019.

\bibitem{Lemme2010}
M.~Lemme.
\newblock Current status of graphene transistors.
\newblock 156:499--509, 1 2010.

\bibitem{Liao2010}
L.~Liao, J.~Bai, R.~Cheng, Y.-C. Lin, S.~Jiang, Y.~Qu, Y.~Huang, and X.~Duan.
\newblock Sub-100 nm channel length graphene transistors.
\newblock {\em Nano Letters}, 10(10):3952--3956, 10 2010.

\bibitem{lin2019true}
J.~Lin, Y.~Wang, Q.~Cao, J.~Kuang, and L.~Wang.
\newblock True random number generation based on arrival time and position of
  dark counts in a multichannel silicon photomultiplier.
\newblock {\em Review of Scientific Instruments}, 90(11):114704, 2019.

\bibitem{marangon2017source}
D.~G. Marangon, G.~Vallone, and P.~Villoresi.
\newblock Source-device-independent ultrafast quantum random number generation.
\newblock {\em Physical review letters}, 118(6):060503, 2017.

\bibitem{Matthew2016}
S.~K. {Mathew}, D.~{Johnston}, S.~{Satpathy}, V.~{Suresh}, P.~{Newman}, M.~A.
  {Anders}, H.~{Kaul}, A.~{Agarwal}, S.~K. {Hsu}, G.~{Chen}, and R.~K.
  {Krishnamurthy}.
\newblock $\mu $ rng: A 300–950 mv, 323 gbps/w all-digital full-entropy true
  random number generator in 14 nm finfet cmos.
\newblock {\em IEEE Journal of Solid-State Circuits}, 51(7):1695--1704, 2016.

\bibitem{MAX11300Datasheet}
Maxim Integrated.
\newblock {\em PIXI, 20-Port Programmable Mixed-Signal I/O with 12-Bit ADC,
  12-Bit DAC, Analog Switches, and GPIO}, 2016.
\newblock Rev. 3.

\bibitem{meech2020efficient}
J.~T. Meech and P.~Stanley-Marbell.
\newblock Efficient programmable random variate generation accelerator from
  sensor noise.
\newblock 2020.
\newblock arXiv:2001.05400.

\bibitem{Meric2008}
I.~Meric, M.~Y. Han, A.~F. Young, B.~Ozyilmaz, P.~Kim, and K.~L. Shepard.
\newblock Current saturation in zero-bandgap, top-gated graphene field-effect
  transistors.
\newblock {\em Nature Nanotechnology}, 3(11):654--659, 2008.

\bibitem{metropolis1949monte}
N.~Metropolis and S.~Ulam.
\newblock The monte carlo method.
\newblock {\em Journal of the American statistical association},
  44(247):335--341, 1949.

\bibitem{nguyen2018programmable}
L.~Nguyen, P.~Rehain, Y.~M. Sua, and Y.-P. Huang.
\newblock Programmable quantum random number generator without postprocessing.
\newblock {\em Optics letters}, 43(4):631--634, 2018.

\bibitem{park2019practical}
B.~K. Park, H.~Park, Y.-S. Kim, J.-S. Kang, Y.~Yeom, C.~Ye, S.~Moon, and S.-W.
  Han.
\newblock Practical true random number generator using cmos image sensor dark
  noise.
\newblock {\em IEEE Access}, 7:91407--91413, 2019.

\bibitem{perach2019asynchronous}
B.~Perach et~al.
\newblock An asynchronous and low-power true random number generator using
  stt-mtj.
\newblock {\em IEEE Transactions on Very Large Scale Integration (VLSI)
  Systems}, 27(11):2473--2484, 2019.

\bibitem{raffaelli2018homodyne}
F.~Raffaelli et~al.
\newblock A homodyne detector integrated onto a photonic chip for measuring
  quantum states and generating random numbers.
\newblock {\em Quantum Science and Technology}, 3(2):025003, 2018.

\bibitem{Schwierz2010}
F.~Schwierz.
\newblock Graphene transistors.
\newblock {\em Nature Nanotechnology}, 5(7):487--496, 2010.

\bibitem{srinivasan20102}
S.~Srinivasan et~al.
\newblock 2.4 ghz 7mw all-digital pvt-variation tolerant true random number
  generator in 45nm cmos.
\newblock In {\em 2010 Symposium on VLSI Circuits}, pages 203--204. IEEE, 2010.

\bibitem{2016:CSP:2901318.2901347}
P.~Stanley-Marbell, V.~Estellers, and M.~Rinard.
\newblock Crayon: Saving power through shape and color approximation on
  next-generation displays.
\newblock EuroSys '16, pages 11:1--11:17. ACM, 2016.

\bibitem{chen2017monte}
D.~B. Thomas.
\newblock Acceleration of financial monte-carlo simulations using fpgas.
\newblock In {\em 2010 IEEE Workshop on High Performance Computational
  Finance}, pages 1--6. IEEE, 2010.

\bibitem{thomas2009comparison}
D.~B. Thomas, L.~Howes, and W.~Luk.
\newblock A comparison of cpus, gpus, fpgas, and massively parallel processor
  arrays for random number generation.
\newblock In {\em Proceedings of the ACM/SIGDA international symposium on Field
  programmable gate arrays}, pages 63--72, 2009.

\bibitem{thrun2010toward}
S.~Thrun.
\newblock Toward robotic cars.
\newblock {\em Comm. ACM}, 53(4):99--106, 2010.

\bibitem{tomasi2018model}
A.~Tomasi et~al.
\newblock Model, validation, and characterization of a robust quantum random
  number generator based on photon arrival time comparison.
\newblock {\em Journal of Lightwave Technology}, 36(18):3843--3854, 2018.

\bibitem{wallace2016toward}
K.~Wallace, K.~Moran, E.~Novak, G.~Zhou, and K.~Sun.
\newblock Toward sensor-based random number generation for mobile and iot
  devices.
\newblock {\em IEEE Internet of Things Journal}, 3(6):1189--1201, 2016.

\bibitem{Wang2010}
H.~Wang, Y.~Wu, C.~Cong, J.~Shang, and T.~Yu.
\newblock Hysteresis of electronic transport in graphene transistors.
\newblock {\em ACS Nano}, 4(12):7221--7228, 2010.

\bibitem{xu2019spad}
H.~Xu, N.~Massari, L.~Gasparini, A.~Meneghetti, and A.~Tomasi.
\newblock A spad-based random number generator pixel based on the arrival time
  of photons.
\newblock {\em Integration}, 64:22--28, 2019.

\bibitem{zhang2018architecting}
X.~Zhang, R.~Bashizade, C.~LaBoda, C.~Dwyer, and A.~R. Lebeck.
\newblock Architecting a stochastic computing unit with molecular optical
  devices.
\newblock In {\em 2018 ACM/IEEE 45th Annual International Symposium on Computer
  Architecture (ISCA)}, pages 301--314. IEEE, 2018.

\end{thebibliography}
